\documentclass{aastex63}
\usepackage{graphicx}
\begin{document}

\title{The H$_2$O Spectrum of the Massive Protostar AFGL 2136 IRS 1 from 2 to 13 $\mu$m at High Resolution: Probing the Circumstellar Disk}

\author[0000-0001-8533-6440]{Nick Indriolo}

\affil{ALMA Project, National Astronomical Observatory of Japan, National Institutes of Natural Sciences, 2-21-1 Osawa, Mitaka, Tokyo 181-8588, Japan}
\author[0000-0001-8341-1646]{D. A. Neufeld}
\affil{Department of Physics \& Astronomy, Johns Hopkins University, Baltimore, MD 21218, USA}
\author[0000-0003-4909-2770]{A. G. Barr}
\affil{Leiden Observatory, Leiden University, Leiden, The Netherlands}
\author[0000-0001-9344-0096]{A. C. A. Boogert}
\affil{Institute for Astronomy, University of Hawaii at Manoa, Honolulu, HI, 96822, USA}
\author{C. N. DeWitt}
\affil{USRA, SOFIA, NASA Ames Research Center MS 232-11, Moffett Field, CA 94035, USA}
\author[0000-0001-8913-925X]{A. Karska}
\affil{Institute of Astronomy, Faculty of Physics, Astronomy and Informatics, Nicolaus Copernicus University, Grudziadzka 5, 87-100 Torun, Poland}
\author{E. J. Montiel}
\affil{USRA, SOFIA, NASA Ames Research Center MS 232-11, Moffett Field, CA 94035, USA}
\author[0000-0002-8594-2122]{M. J. Richter}
\affil{Department of Physics, University of California Davis, Davis, CA, 95616, USA}
\author{A. G. G. M. Tielens}
\affil{Leiden Observatory, Leiden University, Leiden, The Netherlands}

\begin{abstract}
We have observed the massive protostar AFGL 2136 IRS 1 in multiple wavelength windows in the near-to-mid-infrared at high ($\sim3$~km~s$^{-1}$) spectral resolution using VLT+CRIRES, SOFIA+EXES, and Gemini North+TEXES. There is an abundance of H$_2$O absorption lines from the $\nu_1$ and $\nu_3$ vibrational bands at 2.7~$\mu$m, from the $\nu_2$ vibrational band at 6.1~$\mu$m, and from pure rotational transitions near 10--13~$\mu$m. Analysis of state-specific column densities derived from the resolved absorption features reveals that an isothermal absorbing slab model is incapable of explaining the relative depths of different absorption features. In particular, the strongest absorption features are much weaker than expected, indicating optical depth effects resulting from the absorbing gas being well-mixed with the warm dust that serves as the ``background'' continuum source at all observed wavelengths. The velocity at which the strongest H$_2$O absorption occurs coincides with the velocity centroid along the minor axis of the compact disk in Keplerian rotation recently observed in H$_2$O emission with ALMA. We postulate that the warm regions of this dust disk dominate the continuum emission at near-to-mid infrared wavelengths, and that H$_2$O and several other molecules observed in absorption are probing this disk.  Absorption line profiles are not symmetric, possibly indicating that the warm dust in the disk that produces the infrared continuum has a non-uniform distribution similar to the substructure observed in 1.3~mm continuum emission.
\end{abstract}

\section{Introduction} \label{sec_intro}

AFGL 2136 IRS 1 (also referred to as CRL~2136, G17.64+0.16, and IRAS 18196$-$1331) is a luminous \citep[$1.0\times10^5$~L$_{\sun}$;][]{lumsden2013} high mass \citep[$45\pm10$~M$_{\sun}$;][]{maud2019} protostar that is believed to be in the latter stages of its evolution due to a variety of observed characteristics \citep[][and references therein]{boonman2003,maud2018}. It is located at a distance of 2.2~kpc away from the Sun \citep{urquhart2014}, and has been extensively observed from centimeter to micron wavelengths, at low and high angular resolution, and low and high spectral resolution. The myriad observations paint a picture where a single, isolated massive protostar is driving a wide angle bipolar outflow through its natal cloud. The large scale outflow is observed in CO emission at millimeter wavelengths, with both the red and blue lobes being about 100\arcsec\ in extent \citep{kastner1994,maud2018}. Closer to the central source (2\arcsec--10\arcsec) the outflow cavity walls are seen in scattered light at near infrared wavelengths \citep{kastner1992,murakawa2008,maud2018}. The cool molecular envelope exhibits ice and dust absorption bands \citep{willner1982,keane2001ice,dartois2002,gibb2004}, as well as molecular emission at millimeter wavelengths \citep{vandertak2000gasgrain,vandertak2000YSO}, but a much warmer component is also inferred from several different molecules seen in absorption in the near-to-mid infrared \citep{mitchell1990co,lahuis2000,keane2001,boonman2003,boonman2003CO2,goto2013_HCl,indriolo2013H2O,goto2019}. The presence of a dust disk on small spatial scales was suggested by near infrared polarization imaging \citep{minchin1991,murakawa2008}, and by mid infrared interferometric observations \citep{dewit2011,boley2013}. A compact source was marginally resolved at centimeter wavelengths along with a cluster of nearby 22~GHz H$_2$O masers \citep{menten2004}, but only with the recent ALMA 1.3~mm continuum observations has  the $93\times71$~mas dust disk been fully resolved \citep{maud2019}. Thermal line emission at 232.687~GHz from the H$_2$O $\nu_2=1$--1, 5$_{5,0}$-6$_{4,3}$ transition has the same spatial extent as the dust emission, and the H$_2$O gas velocities indicate Keplerian rotation within the disk \citep{maud2019}.  It is ideal that the reader has a clear picture of the AFGL~2136 region in mind to best understand the discussion throughout this paper. In particular, Figure 10 of \citet{maud2018} provides an up-to-date schematic diagram of the AFGL~2136 region, and Figures 1 and 2 of \citet{maud2019} present the compact disk observed in dust and gas emission, respectively.


The presence of circumstellar disks around massive protostars has important implications for the formation of high mass stars. The high luminosities ($\gtrsim10^5$~L$_{\sun}$) of massive protostars ionize and evaporate the surrounding molecular cloud, so it is thought that continued accretion onto the central source must be facilitated through disk-like structures \citep[e.g.,][]{krumholz2009,klassen2016}. Spectroscopic observations in the near and mid infrared provide the capability to measure molecular column densities in this accretion disk. By observing transitions out of several different rotational states of the same molecule, it becomes possible to constrain conditions such as the gas temperature, gas density, and radiation field in the gas where that molecule resides. We have previously demonstrated that H$_2$O is an excellent molecule for this purpose due to its large number of strong absorption lines at near-to-mid infrared wavelengths observed in massive protostars \citep{indriolo2013H2O,indriolo2015exes}. Here, we present the first analysis of H$_2$O absorption at high spectral resolution that combines observations from near-to-mid infrared wavelengths toward AFGL~2136~IRS~1. Such an analysis is instructive for future observations with the James Webb Space Telescope, where spectrally unresolved H$_2$O absorption will be detected throughout the infrared.

\section{Observations}

In order to perform high resolution spectroscopy from the near to mid-infrared we utilized three different instrument+telescope combinations: the Cryogenic High-resolution Infrared Echelle Spectrograph \citep[CRIRES;][]{kaufl2004} on UT1 at the Very Large Telescope provided coverage near 2.5~$\mu$m; the Echelon-Cross-Echelle Spectrograph \citep[EXES;][]{richter2018} on board the Stratospheric Observatory for Infrared Astronomy \citep[SOFIA;][]{young2012} provided coverage near 6~$\mu$m; and the Texas Echelon Cross Echelle Spectrograph \citep[TEXES;][]{lacy2002} at Gemini North provided coverage near 12~$\mu$m. A log of the observations is presented in Table \ref{tbl_obslog}, and some further details regarding the execution at each observatory are given here.

\subsection{CRIRES Observations}
CRIRES observations targeting the $\nu_1$ and $\nu_3$ ro-vibrational bands of H$_2$O were made at two reference wavelengths: 2480.0~nm and 2502.8~nm. Due to the lack of a bright natural guide star near AFGL~2136, the adaptive optics system was not utilized. The 0\farcs2 slit was employed, providing a spectral resolving power of about $10^5$, corresponding to $\sim3$~km~s$^{-1}$ resolution. The slit was oriented at a position angle of 45$^{\circ}$ (along a northeast-to-southwest axis) to minimize contamination from the near-IR scattered light observed in the region \citep[e.g.,][]{murakawa2008}. Spectra were obtained in an ABBA pattern along the slit with 10\arcsec\ between the two nod positions and $\pm3$\arcsec\ jitter width. All observations of AFGL~2136~IRS~1 were immediately preceeded by observations of the bright A2IV star HR~6378 using the same setup for use as a telluric standard.

\subsection{EXES Observations}
EXES observations targeting the $\nu_2$ ro-vibrational band of H$_2$O were made in cross-dispersed high-resolution mode targeting central wavenumbers of 1485.24~cm$^{-1}$, 1639.29~cm$^{-1}$, and 1747.25~cm$^{-1}$ (hereafter referred to as the 6.7~$\mu$m, 6.1~$\mu$m, and 5.7~$\mu$m spectra, respectively). The entrance slit had a width of 1\farcs84, providing a resolving power (resolution) of $\sim85$,000 (3.5~km~s$^{-1}$), and a length of about 10\arcsec\ (varies slightly between settings). To facilitate the removal of telluric emission lines, exposures alternated between on-target and a blank sky position 15\arcsec\ away. AFGL 2136 was one of six science targets observed as part of SOFIA program 04-0120 over multiple flights. The bright A1V star $\alpha$~CMa (Sirius) was observed only once at the same three settings given above for use as a telluric standard for our entire program, so it was not necessarily observed on the same night as AFGL~2136~IRS~1. As will be shown here---and in the subsequent paper discussing the other sources---this strategy proved sufficient for removing telluric features from the spectra.

\subsection{TEXES Observations}
TEXES observations targeting pure rotational transitions of H$_2$O (and ro-vibrational transitions of other molecular species) were made in high-med mode (cross-dispersed echelon + 32 l/mm echelle) at reference wavenumbers of 931.5~cm$^{-1}$, 856.0~cm$^{-1}$, 806.5~cm$^{-1}$, and 768.5~cm$^{-1}$. A resolving power (resolution) of $\sim$85,000 (3.5~km~s$^{-1}$) was achieved using the 0\farcs54 wide slit. The target was nodded 1\farcs6 within the $\sim4$\arcsec\ long slit between exposures to facilitate the removal of sky emission lines. Immediately before or after observations of AFGL 2136 IRS 1 the asteroid 16 Psyche was observed using the same configurations as above for use as a telluric standard.

\section{Data Reduction}

\subsection{CRIRES Data Reduction}
The following reduction procedure was performed independently for each of the four CRIRES detectors on each of the six nights that observations occurred. Raw images were processed using the CRIRES pipeline version 2.3.3. Standard calibration techniques, including subtraction of dark frames, division by flat fields, interpolation over bad pixels, and correction for detector non-linearity effects, were applied. Consecutive A and B nod position images were subtracted from each other to remove sky emission features, and all images from each nod position were combined to create average A and B images. Spectra were extracted from these images using the \texttt{apall} routine in \textsc{iraf}\footnote{http://iraf.noao.edu/} and then imported to IGOR Pro.\footnote{https://www.wavemetrics.com} Wavelength calibration was performed using atmospheric absorption lines, and is accurate to $\pm1$~km~s$^{-1}$. Spectra from the A and B nod positions were then averaged onto a common wavelength scale, and were further processed using the shared reduction steps described in Section \ref{subsec_share}.

\subsection{EXES Data Reduction}

Data were processed using the {\tt Redux} pipeline \citep{clarke2015} with the {\tt fspextool} software package---a modification of the Spextool package \citep{cushing2004}---which performs source profile construction, extraction and background aperture definition, optimal extraction, and wavelength calibration for EXES data. We used this software to produce wavelength calibrated spectra for each individual order of the echellogram. These individual spectra were then stitched together using an average of both orders in the overlap regions to produce a continuous spectrum for each of the three separate observations. Further processing is described in Section \ref{subsec_share}.

\subsection{TEXES Data Reduction}
Data were processed using the TEXES pipeline \citep{lacy2002}. The pipeline performs standard calibration techniques, including flat-field calibration, removal of signal spikes, interpolation over dead and/or noisy pixels, subtraction of nod pair images, summation of differenced spectral images, spectral extraction, and wavelength calibration. Output products are wavelength-calibrated spectra for each echelle order. These individual echelle order spectra were then combined onto a common wavelength scale to create a single spectrum for each setting. The 856.0~cm$^{-1}$, 806.5~cm$^{-1}$, and 768.5~cm$^{-1}$ settings have gaps in wavelength coverage between adjacent echelle orders, so the intensities from the individual orders are unchanged by this combination. The 931.5~cm$^{-1}$ setting produces echelle orders that overlap in wavelength, so an average of both orders in the overlap regions was again used to produce a continuous spectrum. Further processing is described in Section \ref{subsec_share}.

\subsection{Shared Data Reduction Steps \label{subsec_share}}
Once data were processed to the point where each entry in Table \ref{tbl_obslog} had a single, wavelength calibrated spectrum, subsequent data reduction methods for all spectra were shared, regardless of the instrument used.
To remove baseline fluctuations and atmospheric features the science target spectra were divided by the corresponding telluric standard spectra using custom macros developed in IGOR Pro that allow for stretching and shifting of the telluric standard spectrum in the wavelength axis, and scaling of the telluric standard intensity according to Beer's law \citep{mccall2001}. The resulting ratioed spectra were then divided by a 30 pixel boxcar average of the continuum level (interpolated across absorption lines) to remove residual fluctuations and produce normalized spectra. Regions where strong atmospheric absorption features result in no useful information were removed from the spectra to improve visualization. Because Earth's orbital motion causes astrophysical lines to shift with respect to atmospheric lines (i.e., observed wavelength) throughout the year, wavelength scales for all spectra were converted to the local standard of rest (LSR) frame.

One final step was required for processing the CRIRES observations since data were acquired at the same settings on multiple nights. Where multiple spectra cover the same wavelength range, they were combined using a weighted average (weighted by 1/$\sigma^2$, where $\sigma$ is the standard deviation of the line-free continuum in each spectrum). In this way, all of the individual CRIRES spectra were combined to form a single normalized spectrum. This spectrum and the normalized spectra resulting from the EXES and TEXES observations are presented in Figures \ref{fig_criresfull}--\ref{fig_texesfull}.

\section{Results \& Analysis}

It is apparent from Figures \ref{fig_criresfull}--\ref{fig_texesfull} that AFGL~2136~IRS~1 displays a wealth of absorption features. In addition to the H$_2$O absorption we identify features due to H$_2^{18}$O ($\nu_2=1$--0 band), HF ($\nu=1$--0 band), CO ($\nu=2$--0 band), HCN ($\nu_2=1$--0 band), C$_2$H$_2$ ($\nu_5=1$--0 band), and NH$_3$ ($\nu_2=1$--0 band). We also find three broad emission features due to the Pfund series of atomic hydrogen, and a set of unidentified absorption features near 2.49~$\mu$m that are also observed toward the massive protostar AFGL~4176 (Karska et al., in preparation).

\subsection{H$_2$O Absorption}

More than 100 absorption features are present in these spectra, including many that are blends due to absorption from multiple transitions. Using the temperature and total H$_2$O column density inferred from our previous study of this source \citep{indriolo2013H2O} and assuming local thermodynamic equilibrium (LTE), we predicted the strength of all H$_2$O absorption lines covered by our data. By comparing these predictions to the observed spectra we identified absorption features that are due to individual transitions of H$_2$O (i.e., are not blends of multiple H$_2$O transitions or blends of an H$_2$O transition and that of another species). Starting with some of the stronger $\nu_3$ band features, we fit absorption lines (in optical depth space) with a sum of two gaussian components to determine average fit parameters. The line center velocities ($v_{cent}$) and line widths ($\sigma_v$) determined from these fits were then used as initial guesses for fitting the other unblended absorption features. During the fitting procedure these free parameters were restricted to values similar to the initial guesses. The line center velocities of both components were constrained to remain within $\pm4$~km~s$^{-1}$ of the initial guesses (20--28~km~s$^{-1}$ and 28--36~km~s$^{-1}$ for components 1 and 2, respectively), and line widths for both components were constrained to the range 1~km~s$^{-1}\leq \sigma_{v} \leq 6$~km~s$^{-1}$. In this manner, all of the unblended H$_2$O absorption features were fit with line profiles that are the sum of two gaussian functions in optical depth. Some example fits are shown in Figure \ref{fig_linefits}.  Note that it is unlikely that the two components utilized in the fit correspond to two distinct physical components, as will be discussed in Section \ref{section_lineprofiles}.

Using the best-fit parameters we determined an integrated optical depth ($\int \tau dv$) for each component. Uncertainties were computed using the covariance matrix returned by \texttt{scipy.optimize.curve\_fit}. The column density in the lower rotational state associated with the transition ($N_l$) was calculated via the standard equation for optically thin absorption:
\begin{equation}
N_l=\frac{8\pi}{A_{ul}\lambda^3}\frac{g_{l}}{g_{u}}\int \tau dv,
\label{eq_state_column}
\end{equation}
where $\lambda$ is the transition wavelength, $A_{ul}$ is the transition spontaneous emission coefficient, and $g_l$ and $g_u$ are the statistical weights of the lower and upper states, respectively, including nuclear spin degeneracy (i.e., $g=(2J+1)(2I+1)$, where $I=0$ for para-H$_2$O and $I=1$ for ortho-H$_2$O). Transition data were taken from the HITRAN database \citep{hitran2012}. Total column densities in each rotational state are taken to be the sum of the column densities determined from the two components, and the resulting $\ln(N_{l}/g_{l})$ are presented in Table \ref{tbl_results}.

\subsection{H$_2$O Rotation Diagrams}

Column densities determined for individual rotational states are used to construct a rotation (Boltzmann) diagram for H$_2$O (Figures \ref{fig_colorcoded_rotationdiagrams} and \ref{fig_einAcoded_rotationdiagram}). For gas in LTE at a single temperature ($T$) the points should form a straight line with a slope equal to $-T^{-1}$. While this is reasonably true in Figure \ref{fig_colorcoded_rotationdiagrams} panel (a), the large scatter in $\ln{(N_l/g_l)}$ is not explained by this simple isothermal absorbing slab model. In particular, the column densities of a significant number of states fall below the best fit straight line for the majority of levels by as much as a factor of ten. To investigate this issue, we explore whether or not certain lower-state and/or transition characteristics correlate with states that appear to be ``under-populated''. 

In Figure \ref{fig_colorcoded_rotationdiagrams} the points on the rotation diagram are distinguished by different discrete properties of the lower rotational state or the observed transition. Panel (b) has points separated by the instrument with which each transition was observed. It is clear that points from different instruments behave in different ways, but it is unlikely that this is an effect of the different instruments themselves (see Section \ref{section_understanding}). The SOFIA/EXES data show significant scatter, so that sub-set of data can be used to further investigate the observed behavior. Panel (c) shows EXES data separated by the change in the rotational quantum number of the transition ($\Delta J=1$, 0, and $-$1 are the $R$, $Q$, and $P$ branches, respectively). There does not appear to be any significant correlation between $\Delta J$ and $\ln{(N_l/g_l)}$. Panel (d) shows EXES data separated by the nuclear spin (the ortho and para nuclear spin configurations correspond to $K_a+K_c$ being odd and even, respectively) of the lower state, and again, there is no correlation with $\ln{(N_l/g_l)}$. Although not presented here, the CRIRES data similarly show no correlation between $\ln{(N_l/g_l)}$ and nuclear spin configuration, and indicate an ortho-to-para ratio of 3:1. Panel (e) separates the EXES data by the three different observed wavelength ranges. While there is some distinction between the different points, ``under-populated'' states are not unique to any given wavelength range, and above about 2500~K there is no discrepancy between any of the points. As such, the scatter does not appear to be caused by the three separate EXES observations at different wavelengths.

In Figure \ref{fig_einAcoded_rotationdiagram} the points on the rotation diagram are distinguished by color scales that are used to indicate the spontaneous emission coefficient of the transition from which each column density was determined (panels (a), (c), and (e)), and the transmission level at the point of deepest absorption within each absorption feature from which column densities were determined (panels (b), (d), and (f)). For EXES observations the ``under-populated'' states have column densities determined from stronger absorption features (panel (b)), and for states with $E_{l}<2500$~K the scatter in $\ln{(N_l/g_l)}$ is correlated with $A_{ul}$, with column densities in ``under-populated'' states derived from transitions with larger spontaneous emission coefficients (panel (a)). Hints of these same trends are possibly present in the CRIRES observations shown in panels (c) and (d), although they are within the range of uncertainties for many of the inferred column densities. Panels (e) and (f) show the same plots for all data. It is clear from panel (f) that there is a correlation between the depth of absorption features and how far the column densities derived from those features deviate from their expected values. Only the TEXES data points deviate from this pattern. A more in-depth discussion about what may cause these behaviors follows in Section \ref{section_understanding}.

Despite the scatter present in the rotation diagrams, a linear fit to the data points can still be used to infer the gas temperature and total H$_2$O column density. It is clear from Figure \ref{fig_colorcoded_rotationdiagrams} though that this fit will change based on the subset of data points being used, as is demonstrated by the results presented in Table \ref{tbl_rotdiagcomp}. Given the large scatter in the EXES data and the relatively linear, self-consistent behavior of the CRIRES data, we adopt the results obtained from the CRIRES data alone: $N({\rm H_2O})=(8.25\pm0.95)\times10^{18}$~cm$^{-2}$ and $T=502\pm12$~K. A fit using the 6.7~$\mu$m EXES data alone results in a temperature that is 1.6 times higher than, and a total H$_2$O column density that is 8 times lower than the adopted values. This implies that results derived from different and/or limited wavelength ranges should be viewed with caution \citep[e.g.,][]{indriolo2015exes}, and that future observations should be planned with care to minimize such effects. Table \ref{tbl_rotdiagcomp} also shows that a fit to all of the EXES data does not agree with our adopted values. As the upper envelope of EXES data points is consistent with the CRIRES data though (panel (b) of Figure \ref{fig_colorcoded_rotationdiagrams}), it should still be possible to infer reasonably accurate values of $T$ and $N({\rm H_2O})$ in sources where only EXES observations have been made by restricting the fit to a subset of the EXES data points. For example, limiting the fit to EXES data where column densities were derived from weak absorption features (line center transmission~$>75$\%; $\tau_0<0.29$) produces results in better agreement with the CRIRES results (see Table \ref{tbl_rotdiagcomp}). In future cases where scatter similar to that shown here is seen in rotation diagrams, we recommend restricting the linear fit to only include state-specific column densities determined from weak absorption lines when deriving total column densities and rotation temperatures.

\subsection{Other species}

\subsubsection{Absorption Lines \label{sect_otherabs}}

\begin{figure}[h!]
\epsscale{0.8}
\includegraphics[width=\textwidth]{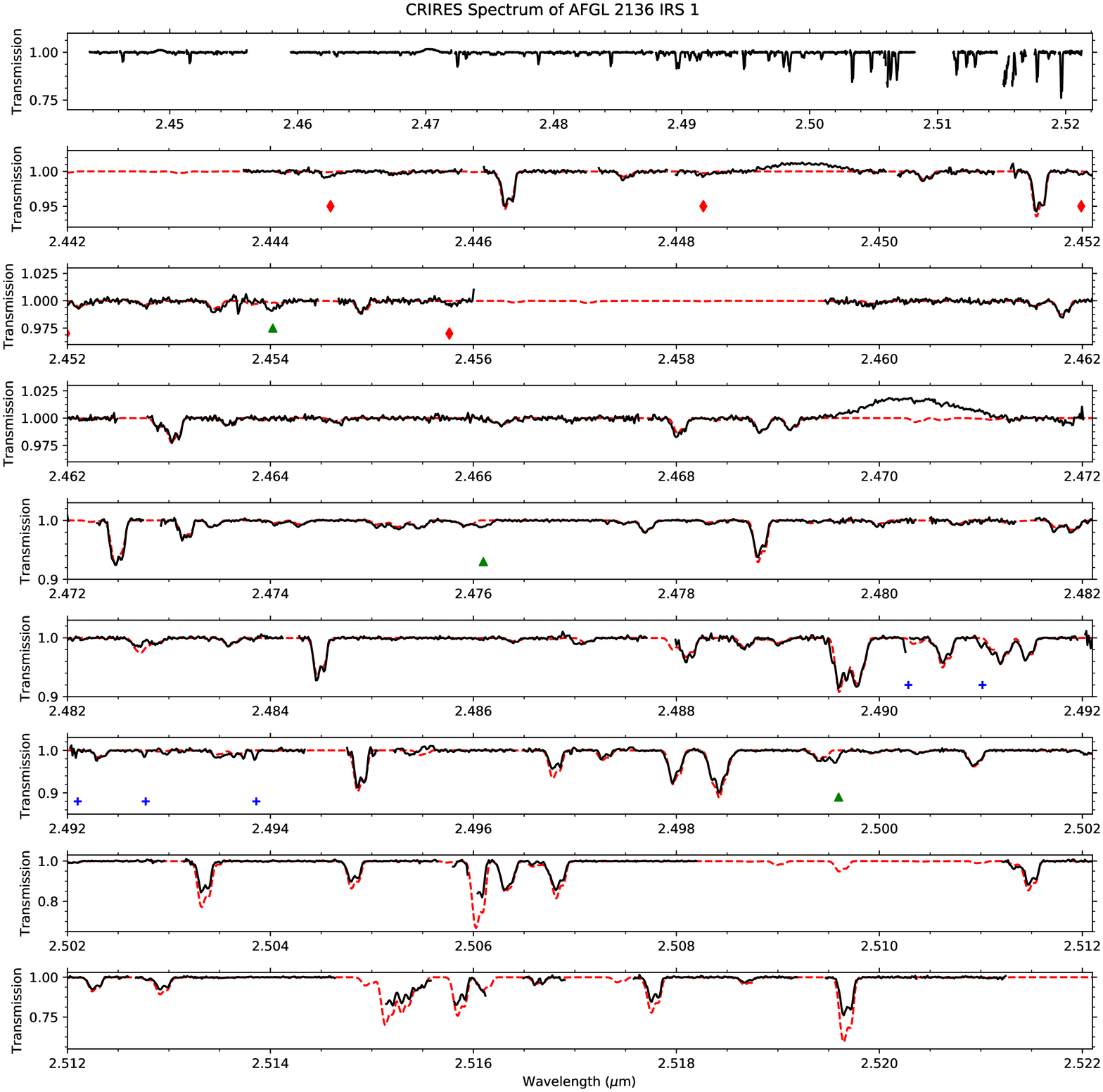}
\caption{Shown here is the near infrared spectrum of AFGL 2136 IRS 1 obtained with CRIRES. The top panel shows the full coverage from about 2.442 $\mu$m to 2.522 $\mu$m, and the remaining panels zoom in on regions that are 0.01 $\mu$m wide. The two largest gaps are due to gaps in wavelength coverage between the CRIRES detectors. The smaller gaps are regions where poor atmospheric transmission precludes any useful analysis. The dashed red curve is the simulated H$_2$O absorption spectrum discussed in Section \ref{sect_otherabs}. Green triangles below the spectra mark HF absorption features, and blue crosses the unidentified features. Red diamonds in the second and third panels mark the locations of the CO $v=2$--0 $P(34)$, $P(35)$, $P(36)$, and $P(37)$ transitions. All remaining absorption features are caused by H$_2$O, while the broad emission features are due to the Pfund series of atomic H. All narrow ``emission'' features are artifacts of the telluric standard division process.}
\label{fig_criresfull}
\end{figure}

\begin{figure}[h!]
\includegraphics[width=\textwidth]{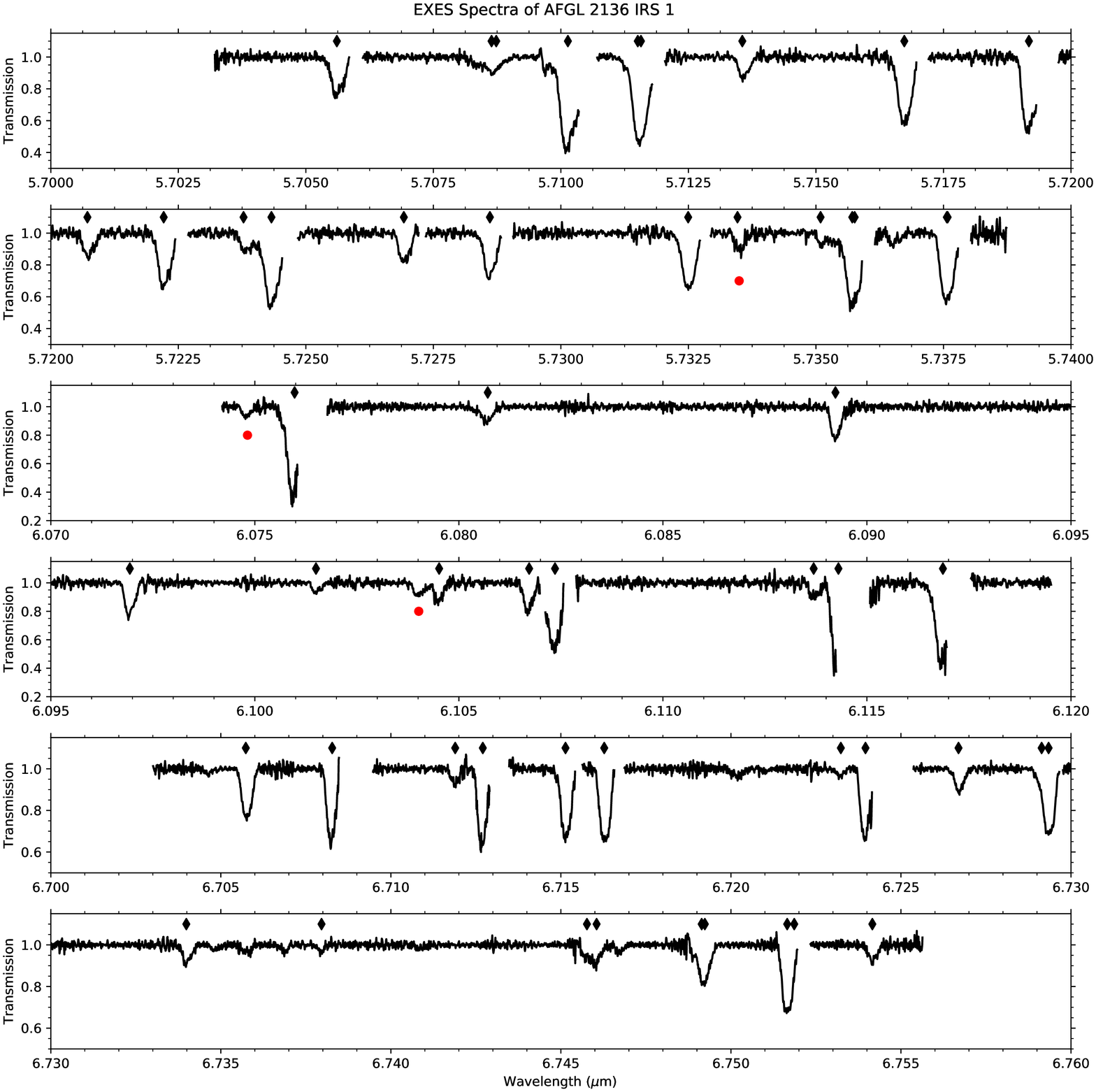}
\caption{Shown here are the mid infrared spectra of AFGL 2136 IRS 1 in select regions from 5.7-6.8 $\mu$m obtained with EXES. Black diamonds above the spectra mark absorption features due to H$_2$O, and red circles beneath the spectra mark features due to H$_2^{18}$O. There are some potentially unidentified features near 6.735~$\mu$m.}
\label{fig_exesfull}
\end{figure}

\begin{figure}[h!]
\includegraphics[width=\textwidth]{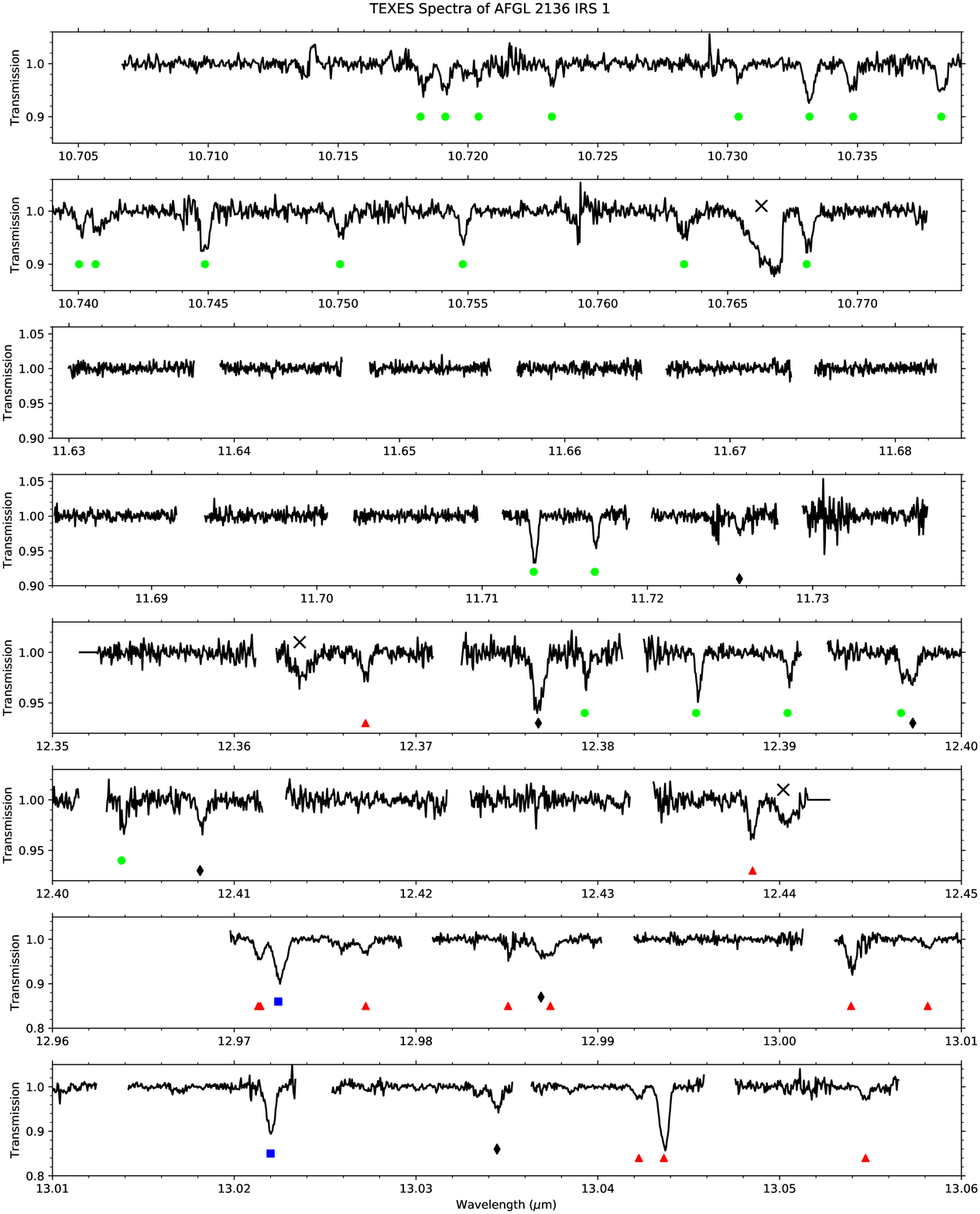}
\caption{Shown here are the mid infrared spectra of AFGL 2136 IRS 1 in select regions from 10.7-13.1 $\mu$m obtained with TEXES. The various symbols beneath the spectra mark absorption features due to NH$_3$ (green circles), H$_2$O (black diamonds), HCN (blue squares), and C$_2$H$_2$ (red triangles). Black crosses above broad features indicate that these are likely artifacts caused by poor matching of the target and telluric standard spectra at the edge of echelle orders.}
\label{fig_texesfull}
\end{figure}

\begin{figure}
\epsscale{0.6}
\plotone{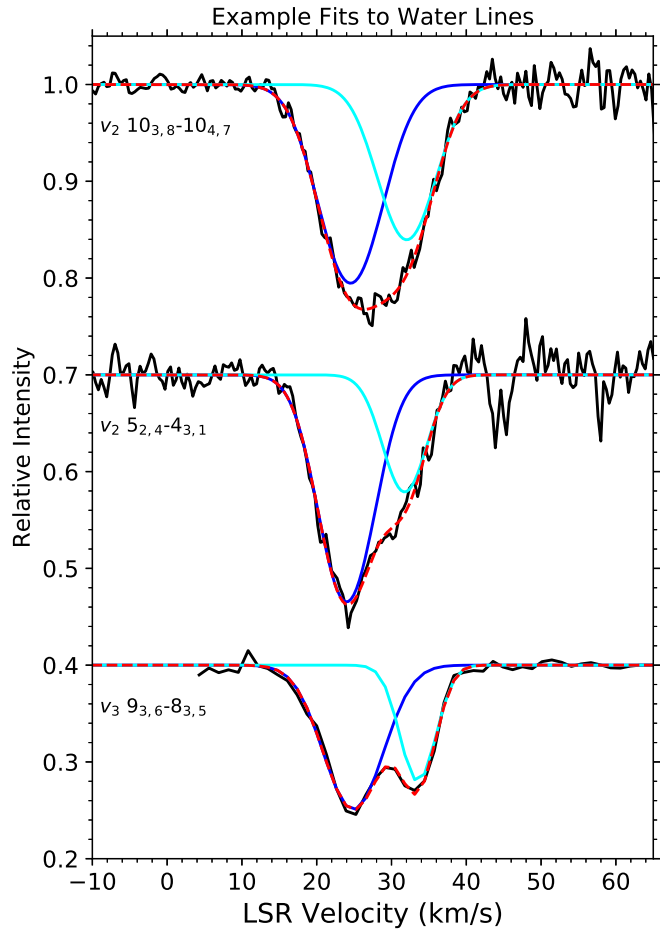}
\caption{Two-component gaussian fits to H$_2$O absorption lines are shown here. The blue and cyan curves mark the components at about 25~km~s$^{-1}$ and 33~km~s$^{-1}$, while the red dashed curve is the sum of both components.}
\label{fig_linefits}
\end{figure}

\begin{figure}
\epsscale{1.0}
\plotone{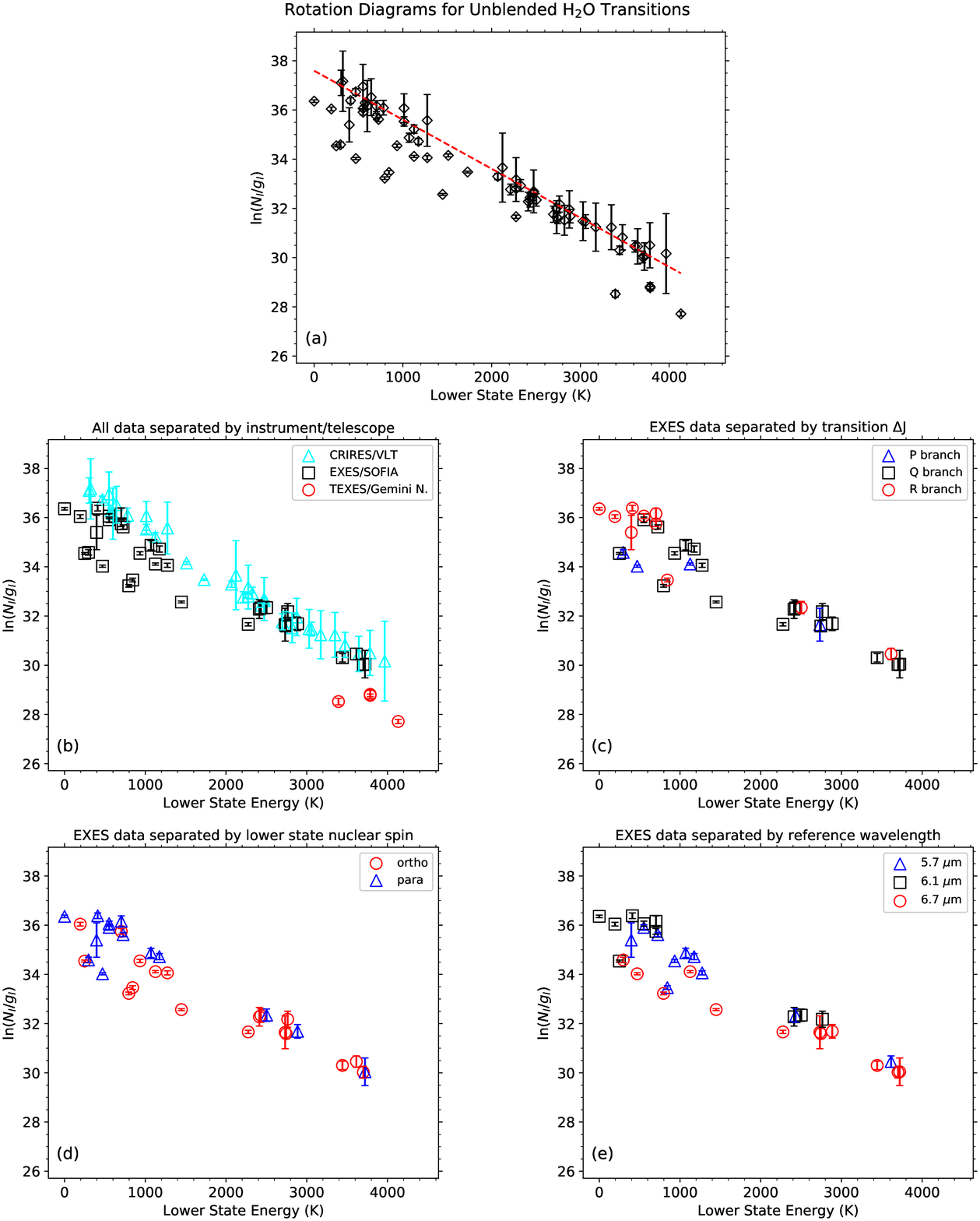}
\caption{These rotation diagrams for H$_2$O in AFGL 2136 IRS 1 are color coded by different discrete properties of the transitions and rotational states from which column densities were determined. Note that for EXES data the uncertainties are generally smaller than the markers. The red dashed line in panel (a) is the best-fit line for the CRIRES data alone, and corresponds to the temperature and H$_2$O column density reported in Table \ref{tbl_rotdiagcomp}.}
\label{fig_colorcoded_rotationdiagrams}
\end{figure}

\begin{figure}
\epsscale{1.0}
\plotone{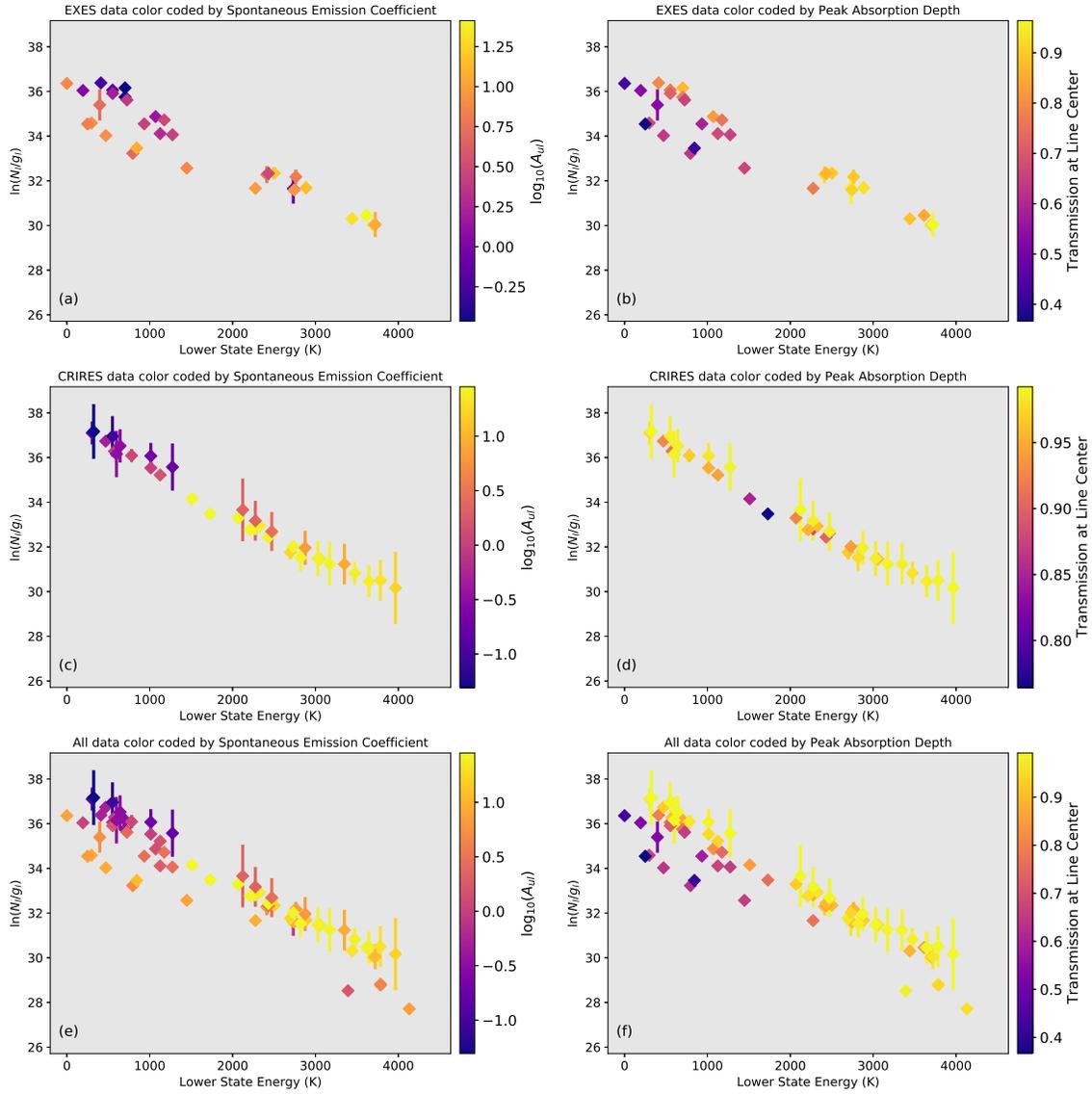}
\caption{These rotation diagrams for H$_2$O in AFGL 2136 IRS 1 are color coded by spontaneous emission coefficient of the transition from which a column density was determined (panels (a), (c), and (e)) and by the maximum depth of the absorption feature from which a column density was determined (panels (b), (d), and (f)). Panels (a) and (b) show only EXES data, panels (c) and (d) show only CRIRES data, and panels (e) and (f) show all data.}
\label{fig_einAcoded_rotationdiagram}
\end{figure}

\begin{figure}
\epsscale{0.6}
\plotone{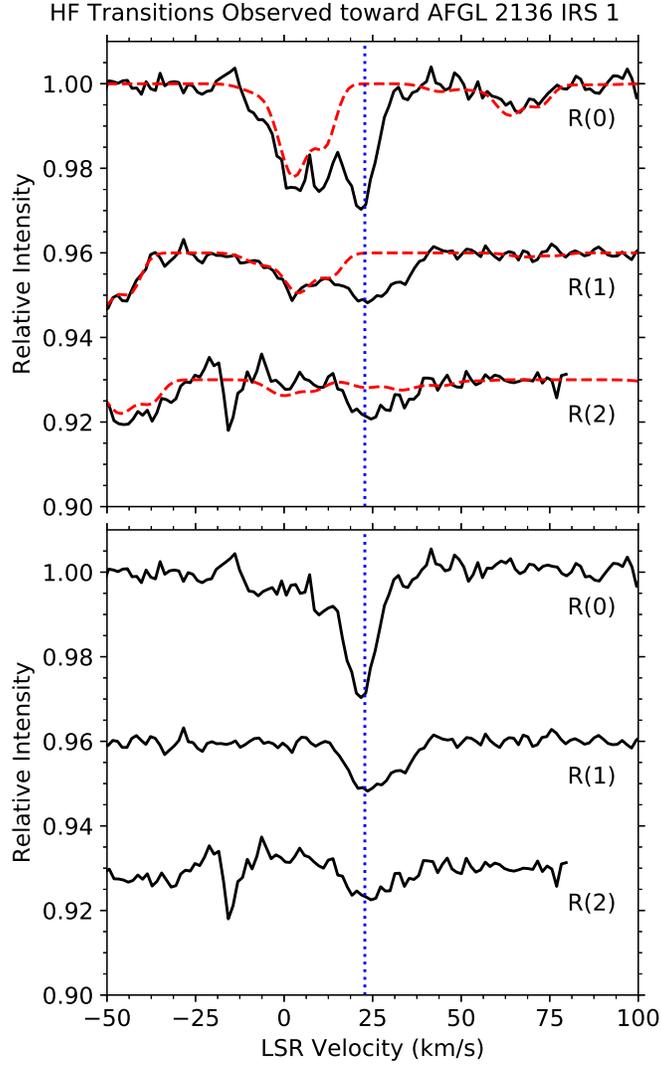}
\caption{Shown here are absorption lines of HF observed toward AFGL 2136 IRS 1. The blue vertical dotted line marks the systemic velocity of 22.8~km~s$^{-1}$ \citep[as measured from molecular emission lines;][]{vandertak2000YSO}. In the top panel solid black lines are the observed spectra and dashed red curves are simulated absorption spectra of H$_2$O. The bottom panels shows the ratio of the observed and simulated spectra to more clearly present the HF line profiles. The narrow feature near $-17$~km~s$^{-1}$ in the $R(2)$ spectrum is due to poor removal of a telluric absorption line.}
\label{fig_HF}
\end{figure}

\begin{figure}
\epsscale{1.0}
\plotone{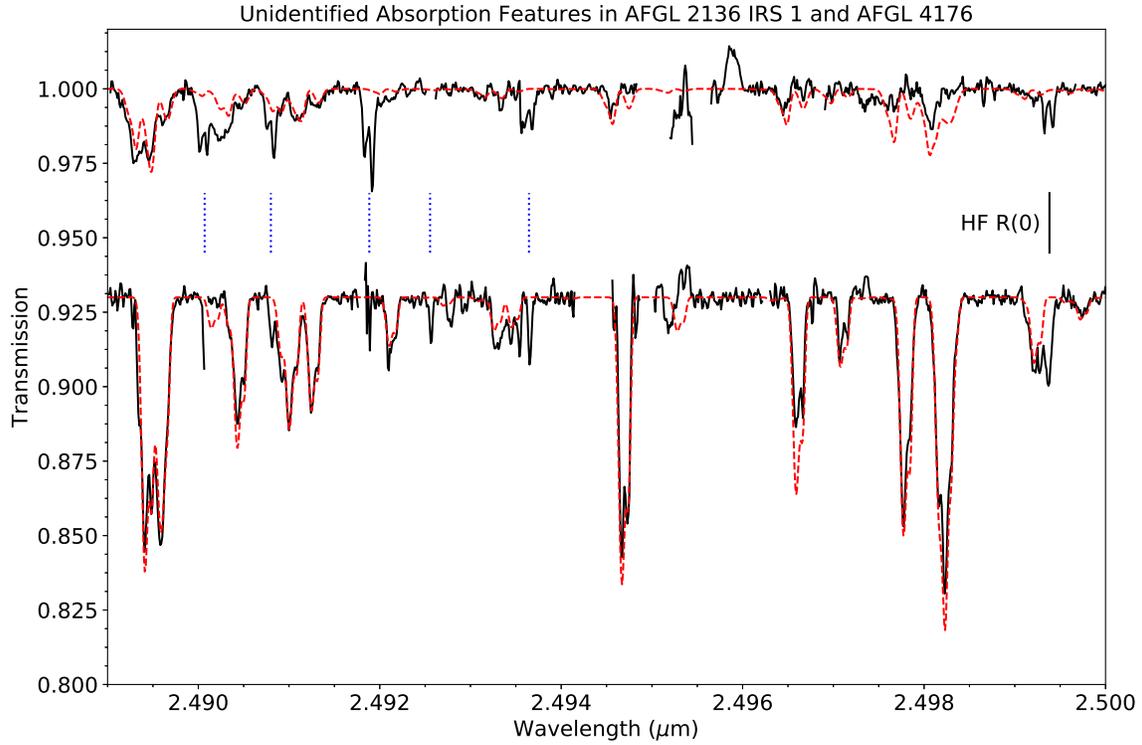}
\caption{Spectra of AFGL 2136 IRS 1 (bottom) and  AFGL 4176 (top; Karska et al., in preparation) highlighting unidentified absorption features are shown here. Dotted blue lines mark the unidentified absorption features at 2.490070~$\mu$m, 2.490800~$\mu$m, 2.491885~$\mu$m, 2.492555~$\mu$m, and 2.493645~$\mu$m. Both spectra have been shifted in wavelength so that absorption lines appear at their rest wavelengths. Red dashed curves are model H$_2$O absorption spectra generated as discussed in Section \ref{sect_otherabs}. The HF $R(0)$ line is marked with a solid black line to demonstrate that its line shape is similar to that of the unidentified features (single peak in AFGL 2136 IRS 1 and double peak in AFGL 4176). This similarity suggests that the species responsible for these absorption lines resides in the cooler foreground gas, not the warm gas where H$_2$O absorption arises.}
\label{fig_crires_ulines}
\end{figure}

A variety of molecular absorption features are observed toward AFGL~2136~IRS~1. Some of these species (NH$_3$, HCN, and C$_2$H$_2$) dominate the absorption features at 10--13~$\mu$m, but at 2.5~$\mu$m H$_2$O is the most prevalent absorber. Despite the fact that a uniform absorbing slab model cannot successfully predict the H$_2$O absorption from 2--13~$\mu$m, such a model is still useful in analyzing the CRIRES observations to search for features caused by other species. Using the average gaussian parameters inferred from the two-component fit described above ($v_{1}=24.8$~km~s$^{-1}$, $\sigma_{1}=4.16$~km~s$^{-1}$, $v_{2}=33.7$~km~s$^{-1}$, $\sigma_{2}=2.92$~km~s$^{-1}$), adopting the total $N({\rm H_2O})$ and $T$ inferred from only the CRIRES data, and assuming that the component at 24.8~km~s$^{-1}$ contains 70\% of the material, we generate a synthetic H$_2$O absorption spectrum (see Figures \ref{fig_criresfull}, \ref{fig_HF}, and \ref{fig_crires_ulines}). The top panel of Figure \ref{fig_HF} shows the synthetic H$_2$O spectrum at the wavelengths of the $v=1$--0 $R(0)$, $R(1)$, and $R(2)$ transitions of HF, demonstrating that all three HF transitions are detected in absorption. The bottom panel of Figure \ref{fig_HF} shows the AFGL~2136~IRS~1 spectrum after division by the synthetic H$_2$O spectrum, effectively showing the absorption due to only HF. By fitting the HF absorption features after division by the synthetic H$_2$O spectrum, we find individual state column densities of $N(0)=(3.1\pm0.8)\times10^{14}$~cm$^{-2}$, $N(1)=(2.6\pm1.3)\times10^{14}$~cm$^{-2}$, and $N(2)=(1.6\pm1.4)\times10^{14}$~cm$^{-2}$, consistent with our previous observations \citep{indriolo2013HF}. The $R(0)$ line shows a different profile from the $R(1)$ and $R(2)$ lines, suggesting that the excited states may be tracing a warmer gas component than the ground state. While less prominent than the HF features, absorption lines due to the $P(34)$--$P(37)$ transitions of the $v=2$--0 band of CO are also detected (Figure \ref{fig_criresfull}). These features are too weak to perform meaningful fits, but their depths are in agreement with predictions based on the temperature and column density of the warm CO component reported by \citet{goto2019}.

Figure \ref{fig_crires_ulines} shows portions of the spectra of AFGL 2136 IRS 1 and AFGL 4176 (Karska et al., in preparation) where the synthetic H$_2$O spectrum fails to reproduce five different absorption features.  To determine the rest wavelengths of these features we matched the line profiles for the unidentified lines with that of the HF $R(0)$ line in the case of AFGL~4176, and that of the H$_3^+$ $R(1,1)^u$ line \citep[from][]{goto2019} in the case of AFGL~2136~IRS~1. The resulting rest wavelengths are 2.490070~$\mu$m, 2.490800~$\mu$m, 2.491885~$\mu$m, 2.492555~$\mu$m, and 2.493645~$\mu$m, with uncertainties of about $5\times10^{-6}$~$\mu$m. We were unable to identify the one or more species that causes these absorption features, although based on the line profiles the carrier must reside in the cooler foreground component, and not the warm gas that gives rise to H$_2$O absorption. 

\begin{figure}
\epsscale{1.0}
\plotone{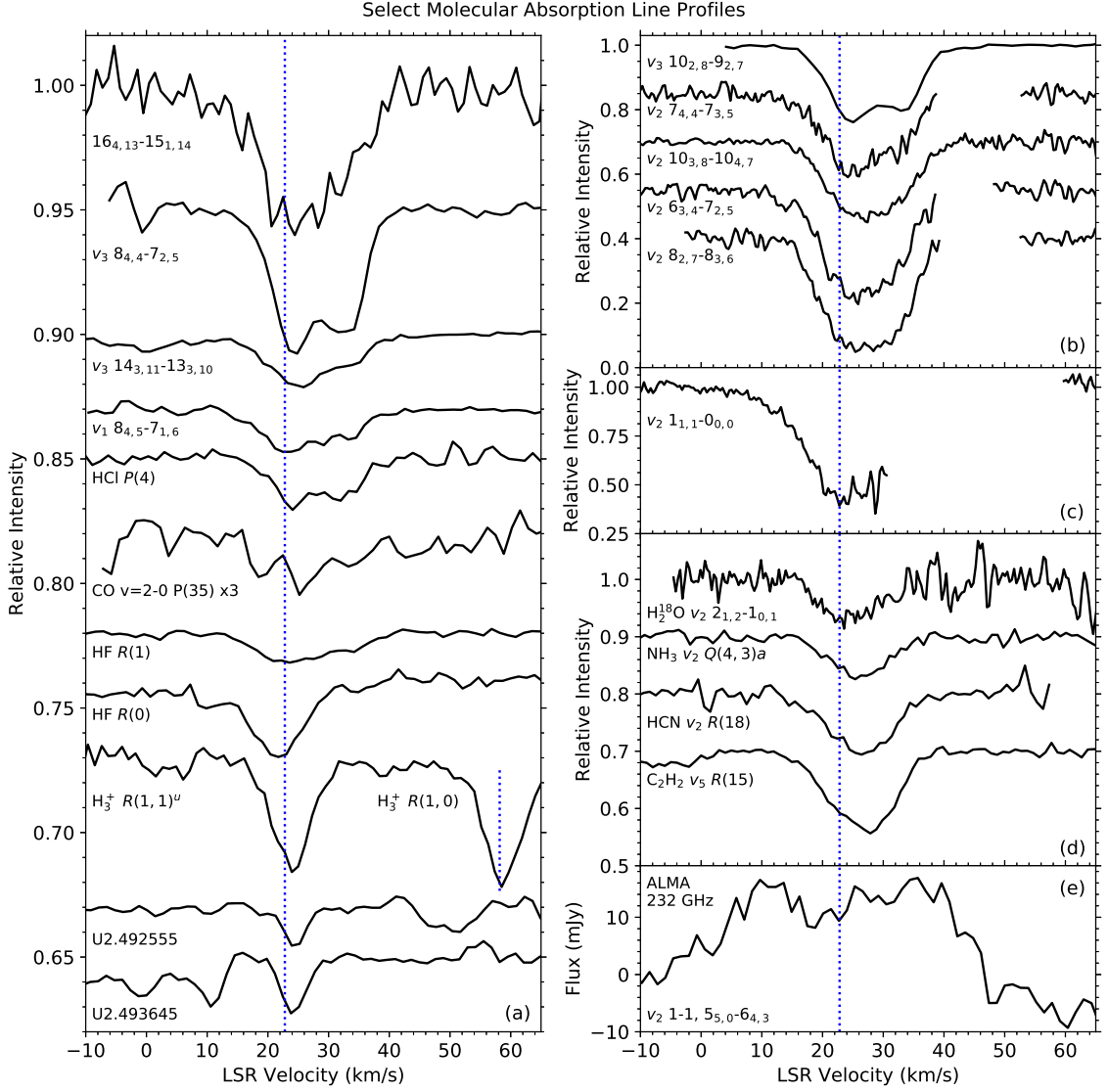}
\caption{A selection of absorption line profiles from H$_2$O and other species observed within our spectra is shown here. Note that instrumental line profiles have not been removed. Labels that do not include a molecule name indicate features due to H$_2$O. Panel (a)  shows weak ($\lesssim5$\% deep) absorption features due to a pure rotational H$_2$O line, lines from the $\nu_1$ and $\nu_3$ vibrational bands of H$_2$O at 2.7~$\mu$m, HCl \citep{goto2013_HCl}, CO (scaled up by a factor of 3), HF (after division by the model H$_2$O spectrum), H$_3^+$ \citep{goto2019}, and two unidentified lines. Panel (b) shows one H$_2$O transition from the $\nu_3$ band, and several from the $\nu_2$ band. Panel (c) shows the $\nu_2$ $1_{1,1}$--$0_{0,0}$ transition of H$_2$O, which probes the ground rotational state. Panel (d) shows absorption from H$_2^{18}$O, NH$_3$, HCN, and C$_2$H$_2$. Panel (e) shows the H$_2$O $\nu_2=1-1$ $5_{5,0}$--$6_{4,3}$ line at 232.687 GHz in emission observed with ALMA \citep[extracted from a rectangular region defined by ($\alpha_{2000}$,$\delta_{2000}$)=($18^h22^m26^s.38857$, $-13^{\circ}30\arcmin12\farcs0226$) and ($18^h22^m26^s.38372$, $-13^{\circ}30\arcmin11\farcs9216$) from the publicly available data cube provided by][]{maud2019}. Inspection of the data cube suggests that line emission ranges from about 3~km~s$^{-1}$ to 45~km~s$^{-1}$. Vertical dashed lines mark the systemic velocity at 22.8~km~s$^{-1}$. In general, there appear to be three categories of line profiles: H$_2$O, HCl, HF $R(1)$, CO $P(35)$, NH$_3$, HCN, and C$_2$H$_2$ show broad, potentially 2-component features that are red-shifted from systemic; H$_3^+$ and the unidentified lines show narrow features that are slightly red-shifted from systemic (with the caveat that rest wavelengths for the unidentified features were determined in part by profile matching to the H$_3^+$ line); HF $R(0)$ shows one component that is at the systemic velocity, and agrees with low-$J$ CO $v=2$--0 absorption profiles \citep{goto2019}. H$_2$O absorption from the ground rotational state extends to lower velocities compared to other transitions, and may also be probing cooler foreground material.}
\label{fig_profiles}
\end{figure}

Example line profiles from all detected species are presented in Figure \ref{fig_profiles}. A variety of strong H$_2$O absorption lines are shown in panel (b), while weak H$_2$O lines are in the top half of panel (a). Panel (c) shows absorption out of the H$_2$O ground rotational state. Panel (d) shows absorption due to ro-vibrational transitions of H$_2^{18}$O, NH$_3$, HCN, and C$_2$H$_2$, all of which have line profiles that are consistent with those observed for H$_2$O in the mid-IR. The bottom half of panel (a) shows absorption due to HCl \citep{goto2013_HCl}, CO, H$_3^+$ \citep{goto2019}, and HF, as well as two unidentified lines. The HCl $P(4)$ absorption profile matches that of H$_2$O, while the profiles of H$_3^+$ and the unidentified lines do not. HF is more complicated as the $R(0)$ line shows one component at the systemic velocity, while the $R(1)$ and $R(2)$ line profiles resemble the H$_2$O absorption. This suggests that the excited HF is tracing the same material as the H$_2$O, while the ground state HF is tracing a cooler foreground component that has been observed at $T\approx60$~K in the low $J$ transitions of the $v=2$--0 band of CO \citep{goto2019}. The NH$_3$, HCN, and C$_2$H$_2$ line profiles also suggest that these species reside in the warm gas component that gives rise to H$_2$O absorption.\footnote{A full analysis of NH$_3$, HCN, and C$_2$H$_2$ in AFGL 2136 IRS 1 using a larger wavelength coverage than presented herein is currently underway (Barr et al. in preparation)} For comparison with the absorption lines, panel (e) shows the H$_2$O $\nu_2=1-1$ $5_{5,0}$--$6_{4,3}$ line at 232.687 GHz observed in emission by ALMA \citep[extracted from the publicly available data cube provided by][]{maud2019}.

\subsubsection{Emission Lines}

\begin{figure}
\epsscale{1.0}
\plotone{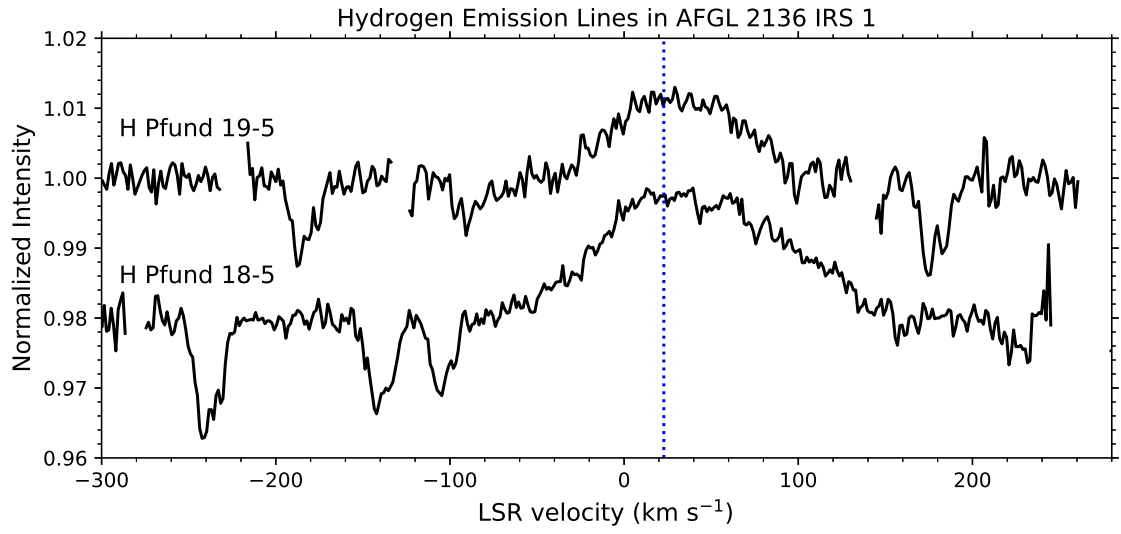}
\caption{These spectra focus on the hydrogen Pfund emission lines observed toward AFGL 2136 IRS 1. The vertical dashed line marks the systemic velocity of 22.8~km~s$^{-1}$.}
\label{fig_HPfund}
\end{figure}

In addition to the plethora of absorption features seen in the AFGL~2136 IRS 1 spectra, a few broad emission features are also present. These can be seen in Figure \ref{fig_criresfull}, and are due to the $n=19$--5 and 18--5 transitions of the Pfund series of atomic H. The 17--5 transition was also seen in emission but suffered heavy interference from telluric H$_2$O lines and so was removed from the spectrum during the normalization process. A zoom in on the 19--5 and 18--5 emission features is shown in Figure \ref{fig_HPfund}. The 19--5 line is centered at 31~km~s$^{-1}$ with a FWHM of 96~km~s$^{-1}$, while the 18--5 line is centered at 39~km~s$^{-1}$ with a FWHM of 127~km~s$^{-1}$. Note, however, that our normalization and baseline removal procedures were optimized for the analysis of narrow absorption features, not broad emission features, so these values are highly uncertain. 

\subsection{Multi-Epoch Observations}

\begin{figure}
\epsscale{0.8}
\plotone{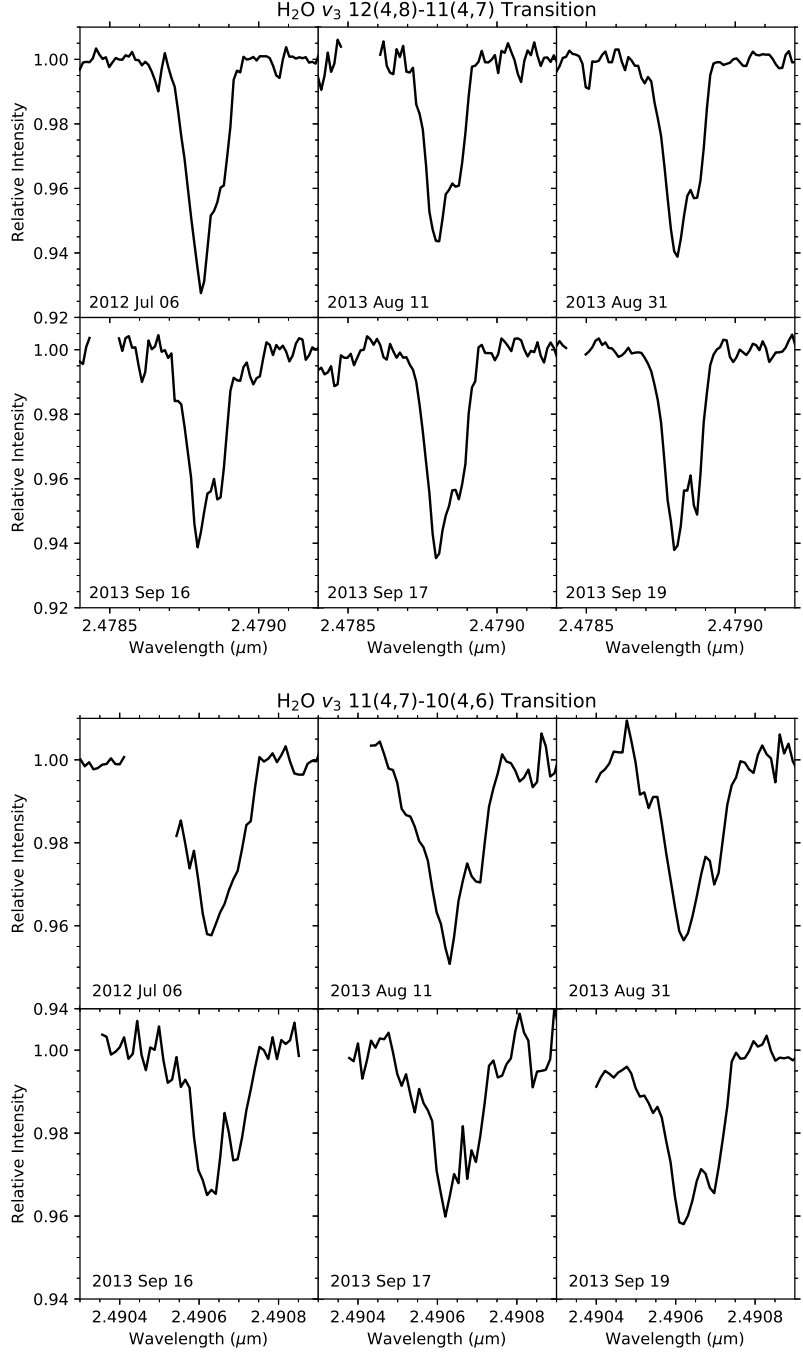}
\caption{Different panels show the absorption profiles of the H$_2$O $\nu_3$ 12$_{4,8}$--11$_{4,7}$ transition (top) and H$_2$O $\nu_3$ 11$_{4,7}$--10$_{4,6}$ transition (bottom) from the six different nights when our VLT/CRIRES observations executed. All spectra have been shifted into the LSR frame.}
\label{fig_multiepoch}
\end{figure}

The CRIRES spectrum presented in Figure \ref{fig_criresfull} is the average spectrum from observations taken on six different nights (see Table \ref{tbl_obslog}). Some unblended H$_2$O absorption lines were covered at all six epochs, so we can check for variability in the absorption profiles. Figure \ref{fig_multiepoch} shows the $\nu_3$ 12$_{4,8}$--11$_{4,7}$ (top panel) and $\nu_3$ 11$_{4,7}$--10$_{4,6}$ (bottom panel) transitions of H$_2$O from each individual night of observations. In the top panel all of the observations from 2013 (39 days maximum separation) agree within the noise level, but the observation from 2012 suggests a deeper absorption feature with a less well-defined shoulder component toward longer wavelengths. In the bottom panel all of the observations, including that from 2012, agree with each other within the noise level. However, this line in 2012 again lacks the well-defined shoulder that is present in all of the 2013 observations. It is difficult to draw any firm conclusions from these spectra.
While the TEXES and EXES observations provide additional epochs in 2014 and 2016, respectively, due to the different instruments utilized, aperture sizes, and transitions observed any such comparison would be highly uncertain. Still, the general agreement of the H$_2$O line profiles from the different telescopes in Figure \ref{fig_profiles} is indicative of no significant changes over a 4 year timescale.

\section{Discussion}

The AFGL~2136 region has been extensively observed over the past few decades at several wavelengths, angular resolutions, and spectral resolutions, with each new study contributing information in the quest to understand its physical and chemical structure. While part of the story certainly unfolds at larger spatial scales---bipolar molecular outflows \citep[100\arcsec;][]{kastner1994,maud2015}, near-IR reflection nebula \citep[10\arcsec; ][]{kastner1992,murakawa2008}, ice absorption in the cold envelope \citep[e.g.,][]{keane2001ice,dartois2002,gibb2004}---here we focus the discussion primarily on the smaller spatial scales immediately surrounding IRS~1.

\subsection{Warm and Hot Molecular Gas}

The first indication of hot molecular gas near AFGL~2136~IRS~1 came from observations of the $v=1$--0 band of $^{13}$CO and $v=2$--1 band of $^{12}$CO in absorption \citep{mitchell1990co}. A hot component was determined to have $N(^{13}{\rm CO})=(2.5\pm0.7)\times10^{17}$~cm$^{-2}$ and $T=580^{+60}_{-50}$~K, containing about twice as much material as a cold component, and was hypothesized to arise in gas close to the central protostar. Recent observations of the $v=2$--0 band of $^{12}$CO confirm this picture, finding $N(^{12}{\rm CO})=(2.8\pm0.4)\times10^{19}$~cm$^{-2}$ and $T=534\pm80$~K \citep{goto2019}. The absorption line profiles for $^{12}$CO $v=2$--0 transitions out of highly excited levels (e.g., $R$(21) and $R$(23)) are consistent with those we find for H$_2$O, suggesting that both species are tracing the same component.

Observations with the {\it Infrared Space Observatory} Short Wavelength Spectrometer ({\it ISO}-SWS) have also revealed absorption from a variety of gas phase molecules tracing warm gas, including CO$_2$, SO$_2$, H$_2$O, HCN, and C$_2$H$_2$. While the large wavelength coverage of {\it ISO}-SWS enables the observation and analysis of entire ro-vibrational bands, the resolution does not allow for the study of kinematics, individual line profiles, and state-specific column densities. Additionally, the determination of total column densities and excitation temperatures requires the use of assumed line widths. Analyses of molecular absorption bands observed toward AFGL 2136 IRS~1 by {\it ISO}-SWS have resulted in a range of excitation temperatures: $T_{ex}({\rm CO}_2)=300\pm100$~K \citep{boonman2003CO2}; $T_{ex}({\rm SO}_2)=350_{-50}^{+100}$~K \citep{keane2001}; $T_{ex}({\rm H_2O})=500_{-150}^{+250}$~K \citep{boonman2003}; $T_{ex}({\rm HCN})=600_{-50}^{+75}$~K \citep{lahuis2000}; $T_{ex}({\rm C_2H_2})=800_{-100}^{+150}$~K \citep{lahuis2000}. The cause of the large scatter in excitation temperatures is unclear, and it is difficult to evaluate if this scatter is real, or if the excitation temperatures agree within uncertainties. Still, it is reasonable to associate these absorption features with the same (or a similar) gas component as that probed by the hot CO. Without kinematic information though, it was not possible to confirm this hypothesis from the {\it ISO}-SWS observations alone.

In the past several years, observations of AFGL~2136~IRS~1 in the near-to-mid infrared have been made at much higher spectral resolution ($\sim3$~km~s$^{-1}$). Our previous observations of H$_2$O with CRIRES confirmed the high excitation temperature ($T_{ex}({\rm H_2O})=506\pm25$~K), but indicated a total H$_2$O column density about seven times larger than that determined from {\it ISO}-SWS \citep[$N({\rm H_2O})=(1.02\pm0.02)\times10^{19}$~cm$^{-2}$ vs. $N({\rm H_2O})=(1.5\pm0.6)\times10^{18}$~cm$^{-2}$;][]{indriolo2013H2O,boonman2003}. The resolved H$_2$O line profiles were in reasonable agreement with the hot CO component from \citet{mitchell1990co}, but without access to the CO data it was still not possible to directly compare the two species. Now, with our new H$_2$O observations and the recent CO $v=2$--0 observations presented by \citet{goto2019}, we can definitively say that both species show the same absorption line profiles, and very likely  reside in the same physical component. This same component may also give rise to the NH$_3$, C$_2$H$_2$, and HCN absorption---as suggested by Figure \ref{fig_profiles}---and this topic will be further addressed by Barr et al. (in preparation). Although HCl shows a similar absorption profile to H$_2$O and CO, the excitation temperature inferred from the observed transitions is significantly lower: $T_{ex}({\rm HCl})\approx250$~K \citep{goto2013_HCl}. This is reminiscent of the scatter in the {\it ISO}-SWS results (if real), and potentially suggests that either: (a) the species do not trace exactly the same material, but perhaps different layers of the same kinematic structure; or (b) excitation temperatures derived from individual molecules are not good indicators of the kinetic temperature of the gas since level populations may be influenced by radiative processes.

ALMA observations of AFGL~2136~IRS~1 have recently revealed emission from the hot molecular gas component as well. The H$_2$O $\nu_2=1$--1, $5_{5,0}$-$6_{4,3}$ transition at 232.687~GHz, which originates from a state $E_u=3461.9$~K above ground, is observed tracing a compact disk ($\sim93$~mas diameter) in Keplerian rotation about a central object with mass $45\pm10$~M$_{\sun}$ \citep{maud2019}. The velocity centroid along the disk minor axis is in agreement with the line-center velocity of the stronger component found from our gaussian fitting procedure (24.8~km~s$^{-1}$), suggesting that the H$_2$O absorption we observe in the IR may arise in the disk now imaged at sub-mm wavelengths. Compact SiO emission ($J=5$--4; 217.105~GHz; $E_u=31.3$~K) has also been observed toward AFGL~2136~IRS~1 with ALMA. A portion of the SiO emission shows similar kinematics as H$_2$O and likely traces the disk \citep{maud2019}, but emission near the systemic velocity is more spatially extended and hypothesized to arise in a rotating disk wind \citep{maud2018}. Perhaps different conditions in these two components are responsible for the different rotation temperatures inferred from various molecules.

\subsection{Hot Ionized Gas}

Even more compact than the disk around AFGL~2136~IRS~1 is a region producing emission in transitions indicative of ionized gas. The H30$\alpha$ recombination line is spatially unresolved with ALMA's 20~mas beam, but there is a hint of rotation in the same sense as the H$_2$O emission \citep{maud2019}. It has a very broad line profile that is centered at 22.1~km~s$^{-1}$, with $FWHM\sim81.9\pm1.7$~km~s$^{-1}$ \citep{maud2018}. This is likely the same region that gives rise to the H Pfund $n=19$--5 and 18--5 emission lines shown in our Figure \ref{fig_HPfund} as the electrons cascade down toward the ground electronic state following recombination. 
Similar emission features from the $n=21$--6 and 19--6 transitions of the Humphreys series of atomic H near 3.6~$\mu$m have also been observed toward AFGL~2136~IRS~1 (M. Goto, private communication), and the hydrogen Br$\gamma$ ($n=7$--4) line has been observed with $FWHM_{\rm Br\gamma}\approx133$~km~s$^{-1}$ \citep{murakawa2013}. Hydrogen Br$\gamma$ emission in massive protostars has previously been interpreted as emission from an outflowing stellar wind or disk wind \citep[e.g.,][]{malbet2007,murakawa2013}, although the spatially unresolved H30$\alpha$ emission requires the emitting region to be much smaller than the circumstellar disk.

\subsection{Understanding the H$_2$O Spectrum \label{section_understanding}}

The optimal use of spectroscopy in studying massive protostars entails understanding what underlying characteristics cause the observed absorption and emission features. While the use of a simple isothermal absorbing slab model that is removed from and fully covering a background source can be instructive, its assumptions and limitations---many of which are described in \citet{lacy2013}---prevent such a model from re-producing observed spectra. To re-iterate some of those points: (1) the central protostar will produce a temperature gradient in the surrounding gas and dust; (2) the absorbing gas can be mixed with the warm dust that serves as the ``background'' continuum; (3) molecules both emit and absorb light. Here, we discuss how these and other effects can influence the observed spectra.

A covering fraction of less than one (i.e., clumpy absorbing material that does not cover the entire background source) is a scenario that has previously been invoked to explain observed IR absorption spectra \citep[e.g.,][]{knez2006,knez2009,barentine2012,indriolo2015exes}. In this case, saturated absorption lines will have a depth equal to the covering fraction, and all lines will appear weaker than expected, since total absorption corresponds to a relative intensity larger than zero. While some absorption features are about 4 times weaker than expected assuming the H$_2$O column density and excitation temperature inferred from CRIRES data alone (e.g., $\nu_2$ 2$_{1,1}$-3$_{2,2}$ at 6.723374~$\mu$m), others are about as strong as expected (e.g., $\nu_2=2-1$ 7$_{1,6}$-7$_{2,5}$ at 6.726109~$\mu$m). Since covering fraction would affect all line depths equally, this explanation does not seem feasible. Even if the absorbing material does not fully cover the background source, the deepest absorption feature in the spectrum (H$_2$O $\nu_2$ 3$_{2,1}$-3$_{1,2}$ at 6.075447~$\mu$m) reaches a transmission level of about 0.3, so a minimum covering fraction of 70\% is required for the EXES observations. 

One of the obvious differences amongst our spectra is the instruments used in performing the observations. In particular the CRIRES, EXES, and TEXES observations were made using slits with different widths and lengths (widths of 0\farcs2, 1\farcs84, and 0\farcs54, respectively), so light from slightly different regions is being observed in each case. The larger apertures could probe regions with different conditions than probed by the narrowest (CRIRES) slit from which we infer column density and temperature. Extended continuum emission (assuming the absorbing gas is not extended) or extended line emission could fill in the absorption features. The extended continuum explanation is equivalent to a covering fraction argument though, and so is ruled out. Extended line emission would have to be limited to only those transitions that are weaker than expected, but should also manifest as emission lines if a spectrum is extracted slightly off of the continuum peak. No H$_2$O line emission is seen when we extract a spectrum slightly off source, so this scenario seems unlikely as well.

Previous observations of molecular absorption and emission lines in the mid-IR have demonstrated different behaviors of transitions within a ro-vibrational band.
The $\nu_2$ band of H$_2$O has shown some lines in emission and others in absorption \citep{gonzalez-alfonso1998,gonzalez-alfonso1999}, as has the $\nu_2$ band of NH$_3$ \citep{barentine2012}. In both of these studies though, there is a pattern that the $R$-branch lines appear in absorption, while the $P$-branch lines appear in emission, explained by the relative importance of the radiative pathways that populate and de-populate the vibrationally excited rotational levels. If this mechanism were responsible for the large scatter in the rotation diagrams---not to the point of causing emission, but just enough to fill in some of the absorption---then there should be a discrepancy in the column densities inferred from $R$-branch and $P$-branch transitions. Panel (c) of Figure \ref{fig_colorcoded_rotationdiagrams} shows no correlation between ``under-populated'' states and transition $\Delta J$ though, so this emission mechanism is also unable to explain the observed spectra.

The parameter that shows the clearest correlation with a state being ``under-populated'' is the depth of the absorption feature from which the state-specific column density was derived (Figure \ref{fig_einAcoded_rotationdiagram}, panels (b), (d), and (f)). Excepting the absorption features observed with TEXES, deeper absorption lines tend to result in more ``under-populated'' states. This can be explained by a scenario where the absorbing gas is well mixed with the warm dust that serves as a ``background'' source of continuum emission. Blackbody continuum emission predominantly comes from the $\tau\approx1$ ``surface'', so any observed line absorption only comes from gas above that surface. The depth of the $\tau\approx1$ ``surface'' depends on the dust opacity (and associated absorption coefficient, $\kappa_\nu$), which is dependent on wavelength \citep{draine2003} and on the distribution of dust grain sizes \citep[e.g.][]{agurto-gangas2019}, so the layers of gas probed by observations at 2~$\mu$m, 6~$\mu$m, and 13~$\mu$m may be different. Because the absorption coefficient is also affected by line processes, $\kappa_\nu$ is larger at the wavelengths of transitions with larger spontaneous emission coefficients. This means that the physical depth where $\tau\approx1$ is shallower for stronger lines than for weaker lines, such that there is a smaller column of gas in front of the ``background'' source. In this case, it is not that any levels are under-populated with respect to LTE, just that they appear that way from a simple absorbing slab analysis since different transitions probe to different depths within the absorbing gas.

The above scenario requires that the absorbing gas and source of continuum emission be well mixed. Mid-IR (8--13~$\mu$m) interferometric observations of AFGL~2136~IRS~1 are suggestive of an elongated structure that has a size and orientation reasonably consistent with the recently reported sub-mm disk \citep{dewit2011,boley2013,maud2019}. This indicates that the continuum emission at wavelengths longer than about 8~$\mu$m likely arises in the disk, and so it is reasonable to assume that the gas is mixed with the emitting dust for the TEXES observations. For silicate dust at the dust sublimation temperature \citep[$T\sim1300$~K;][]{kobayashi2011} the blackbody spectrum peaks at about 2.2~$\mu$m, so the dust disk is expected to be bright in the near infrared as well. Radiative transfer models of near infrared interferometric observations of another massive protostar suggest that the dominant continuum emission source at 2.2~$\mu$m in that system is a circumstellar disk \citep[][IRAS 13481$-$6124]{kraus2010}, so thermal emission from the warm dust disk may be the dominant continuum source in all of our observations from 2.5--13~$\mu$m \citep{beltran2016}. 


Of the effects discussed above, the mixing of warm dust and absorbing gas seems to be the most likely explanation for the observed H$_2$O absorption spectrum. ALMA observations have shown that the dust disk and gas disk have roughly the same physical extent \citep{maud2019}, and it is plausible that the warm dust in the disk dominates the continuum at all observed wavelengths. The H$_2$O absorption line profiles, however, are not consistent with expectations in a scenario where the entire disk is responsible for continuum emission and molecular absorption. To better understand what may cause this discrepancy, a more detailed investigation of the line profiles is required.


\subsection{Line Profiles} \label{section_lineprofiles}

Given the gas temperature inferred for the AFGL~2136~IRS~1 disk, line profiles are expected to be dominated entirely by orbital motions rather than thermal broadening. As such, information about gas kinematics is contained within the absorption and emission line profiles. H$_2$O absorption lines observed with CRIRES show two distinct absorption peaks while those observed with EXES and TEXES do not, but in all cases lines are well described by two gaussian components. The robustness of individual gaussian fit parameters is difficult to assess, as in many cases intensity can be traded between the two components without adversely affecting the overall fit, so it is not particularly instructive to consider the properties of the two fit components separately. Indeed, the single temperature required to describe the rotation diagram for CRIRES data and the relatively constant intensity ratio between the two components in the CRIRES data suggest that all of the absorption arises in a single physical component.

\begin{figure}
\epsscale{1.0}
\plotone{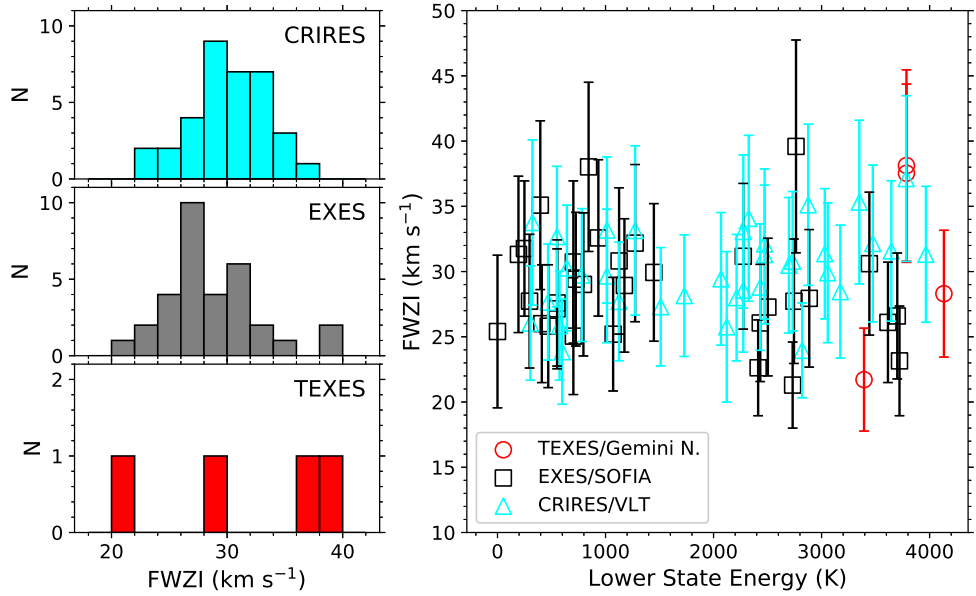}
\caption{The distribution of FWZI line widths ($|v_2-v_1|+3\sigma_1+3\sigma_2$) is plotted for H$_2$O absorption features observed with the three different instruments in the panels on the left hand side. There is a marginal decrease in FWZI from the CRIRES data to the EXES data. The right hand panel shows FWZI as a function of lower state energy for all unblended H$_2$O lines. No correlation is seen between the two parameters. FWZI line width uncertainties are dominated by the fits to the line wings.}
\label{fig_linewidths}
\end{figure}

Absorption lines arising in circumstellar disks have previously been observed toward FU Ori type objects \citep[young, low mass stars experiencing luminosity bursts due to enhanced accretion;][and references therein]{hartmann1996}. The fact that the lines appear in absorption rather than emission is indicative of a temperature inversion in the disk, where heating due to accretion causes the disk midplane to be warmer than the surface. Two properties observed in absorption lines toward FU Ori type objects are that: (1) many absorption features are double-peaked; and (2) the width of absorption features tends to decrease with increasing wavelength. The double-peaked nature of the lines is explained by a disk in Keplerian rotation, with the two absorption peaks caused by the blueshifted and redshifted sides of the disk. While we do see double-peaked absorption lines toward AFGL~2136~IRS~1, the underlying explanation is not quite the same, as will be discussed below. The anticorrelation between linewidth and wavelength is explained by the temperature in the disk decreasing with increased distance from the central source. Continuum emission at shorter wavelengths preferentially arises in warmer material at small radii where orbital velocities are large, so the absorbing gas produces broad features. At longer wavelengths the continuum emission arises from larger radii where orbital velocities are smaller, hence narrower absorption lines. An analysis of the distribution of full-width-at-zero-intensity (FWZI) line widths (defined as $|v_2-v_1|+3\sigma_1+3\sigma_2$) shows a marginal decrease (1.5~km~s$^{-1}$ in the mean value) between H$_2$O lines at 2.5~$\mu$m and 5.7--6.7~$\mu$m (Figure \ref{fig_linewidths}). As FWZI line widths are broad (30~km~s$^{-1}$ on average) and instrumental profiles have not been removed from the spectra, it is difficult to determine if this shift is real.

The line profile for the 232 GHz H$_2$O emission extracted over the full extent of the disk is shown in Figure \ref{fig_profiles} panel (e). Inspection of the data cube channel by channel suggests that the flux at $v\lesssim3$~km~s$^{-1}$ is an artifact of the image processing, and that the H$_2$O emission from the disk extends from about 3~km~s$^{-1}$ to about 45~km~s$^{-1}$. Emission at these extreme velocities arises from the innermost portions of the disk along the major axis, and the velocity centroid along the disk minor axis is halfway between these two extremes at 24~km~s$^{-1}$. The line profile has two emission peaks, with one each arising from the redshifted and blueshifted halves of the disk. All of these features are consistent with a disk in Keplerian rotation, as reported by \citet{maud2019}.

H$_2$O absorption profiles---and those of other molecules---clearly do not show the same characteristics as the 232~GHz H$_2$O emission. The strongest H$_2$O absorption occurs at about 25~km~s$^{-1}$ with a second weaker component at about 33~km~s$^{-1}$, and absorption extends from about 12 to 40~km~s$^{-1}$. The narrower (with respect to emission), asymmetric line profiles seen in absorption suggest that if the H$_2$O absorption arises within the disk, it is not tracing the entire disk. In particular, the absence of absorption at the extreme velocities, and the lack of an absorption component at $v\leq24$~km~s$^{-1}$ indicate that H$_2$O absorption probes neither the innermost portion of the disk along the major axis, nor the blueshifted half of the disk. An examination of Figure 1b in \citet{maud2019} provides a potential explanation for this observed behavior. There, the 1.3~mm continuum emission shows clumpy structure within the disk after an assumed point source caused by free-free emission has been removed. The strongest continuum peaks within the disk are located along the minor axis, where $v\approx24$~km~s$^{-1}$ is expected, and near the outer edge of the redshifted side, where $v\approx33$~km~s$^{-1}$ is expected. These velocities are in good agreement with the two absorption components that are observed. If the clumpy structure seen in the 1.3~mm continuum emission is indicative of the 2--13~$\mu$m continuum structure, this could explain the molecular absorption profiles seen in the near-to-mid-IR.

As the ALMA observations were performed on 2017 Oct 10, the disk sub-structure at the time of the IR observations would have been oriented slightly differently. Material at the inner and outer disk radii of 30~AU and 120~AU has orbital periods of about 24~yr and 196~yr, respectively (assuming a central mass of 45~$M_\sun$), and would move by at most 18~mas and 9~mas in the plane of the sky over the 5.3~yr between our first CRIRES observation and the ALMA observations given a distance of 2.2~kpc. This suggests that the majority of the substructure identified in the ALMA observations was in a reasonably similar orientation at the time when the IR observations were performed.

Finally, we note that there is no evidence for a high velocity wind component in absorption toward AFGL~2136~IRS~1. CO absorption lines observed toward some massive protostars show broad blueshifted wings, extending over $-100$~km~s$^{-1}$ from the systemic velocity \citep[e.g.,][]{vandertak1999}. Neither the H$_2$O nor the CO \citep{mitchell1990co,smith2014,goto2019} absorption lines observed toward AFGL~2136~IRS~1 show such a broad component. Most likely this is simply due to the orientation of the disk and outflow system, such that wind and outflow components are not probed by the pencil beam line of sight toward the continuum source.

\subsection{Conditions in the AFGL~2136~IRS~1 Disk}

In our previous study of H$_2$O absorption observed with CRIRES toward AFGL 2136 IRS 1 we used a statistical equilibrium code to infer conditions within the absorbing gas \citep{indriolo2013H2O}. The code modeled the excitation of  the lowest 120 rotational states of ortho- and para-H$_2$O to determine the expected level populations under a variety of physical conditions. In this code the effects of radiative trapping are treated using an escape probability method, and collisional rate coefficients between H$_2$O and H$_2$ were taken from \citet{daniel2011}, and were extrapolated to higher energy states for which calculations are unavailable using an artificial neural network method \citep{neufeld2010_nn}. Radiative pumping effects in both ro-vibrational and rotational transitions due to the infrared continuum emission from AFGL~2136~IRS~1 were also included, assuming the spectral energy distribution implied by a fit to the data obtained by \citet[][the TO model plotted in their Figure 7]{murakawa2008}. This analysis demonstrated that the observed state specific column densities could be explained by purely collisional excitation, purely radiative excitation, or some contribution from both mechanisms between these two extremes. In the collisionally dominated case the gas is at high density and temperature ($n({\rm H_2})\gtrsim5\times10^9$~cm$^{-3}$; $T=506\pm25$~K). In the radiatively dominated case a $T=700$~K blackbody subtends 2$\pi$~sr (i.e., half of the sky) as seen by the gas, the gas temperature is unconstrained, and the density must be $n({\rm H_2})\lesssim10^6$~cm$^{-3}$. Note that these two scenarios are limiting cases, and that some combination of collisional and radiative excitation is also viable, in which case the inferred constraints are relaxed (e.g., if excitation is primarily through collisions, but also due in some part to radiation, then the lower limit on the density will decrease compared to that in the purely collisional case).

We repeated this same statistical equilibrium analysis on the H$_2$O column densities presented here in Table \ref{tbl_results}, but were unable to find a model capable of simultaneously reproducing all values. This failure can be attributed to the large scatter in the rotation diagrams caused by different transitions probing different depths within the disk, as discussed above. Limiting the analysis to only the CRIRES data produces the same results as in \citet{indriolo2013H2O} that are summarized above, so we do not present the entire analysis anew in this paper.

Given the limits on gas density and the measured H$_2$O column density, we can place constraints on the size of the region that contains H$_2$O. The relative abundance of H$_2$O with respect to H$_2$ in warm gas is expected to be about 10$^{-4}$ \citep[][and references therein]{vandishoeck2013}, so the radiatively dominated case suggests an upper limit of $n({\rm H_2O})\lesssim 100$~cm$^{-3}$. With $N({\rm H_2O})=8.25\times10^{18}$~cm$^{-2}$, this requires a minimum path length of $l\gtrsim8.25\times10^{16}$~cm, or about 5500~AU. The collisionally dominated case requires $n({\rm H_2O})\gtrsim 5\times10^5$~cm$^{-3}$, corresponding to a maximum path length of $l\lesssim1.65\times10^{13}$~cm, or about 1.1~AU. It is clear that the path length over which absorption occurs in the radiatively dominated case is incompatible with the $r\sim120$~AU size scales of the AFGL~2136~IRS~1 disk. The path length in the collisionally dominated case however, is consistent with the picture where the observed H$_2$O absorption is coming from the upper layers of the disk. That said, the fact that the H$_2$O resides within the disk means that blackbody radiation from the dust disk must have some influence on the excitation of H$_2$O, so the absorbing layer can be thicker than that determined for the scenario where only collisions are considered.

Although Figure \ref{fig_multiepoch} shows the possibility of marginal variability in H$_2$O absorption line profiles, the similarity of the line profiles in the CRIRES, EXES, and TEXES data shown in Figure \ref{fig_profiles} suggests no significant changes in the absorbing material over 4 year time scales. This is further evidence against a scenario where the background continuum source and absorbing gas are separated (e.g., gas in the disk absorbing light from the central protostar), because the absorbing material in a pencil beam toward the central object will change as it orbits. The time between our first CRIRES observation and last EXES observation was 1357 days. Over this time period, gas at radii of 30~AU and 120~AU in Keplerian rotation about a 45~$M_{\sun}$ star would move 28.6~AU and 14.3~AU along their respective orbits. The disk material would have to be uniform over these size scales in order for the absorption profile to remain nearly constant in a pencil beam toward the central object over 4 years. If portions of the disk itself serve as the background continuum source though, then the material above the $\tau\approx1$ ``surface'' will be probed by absorption spectroscopy, regardless of where the material is in its orbit. Time variability in line profiles is still expected in this case though, as the 1.3~mm continuum emission shows substructure \citep{maud2019}. As regions of brighter continuum emission orbit the central object, the absorption from the gas associated with these regions will shift in velocity.


\subsection{H$_2$O Isotopologues}
In our observations, absorption due to both the H$_2^{16}$O and H$_2^{18}$O isotopologues of H$_2$O is detected. With only two unblended H$_2^{18}$O lines observed we cannot estimate the total column density of that isotopologue, but we can compare state-specific column densities. Transitions probing the $J_{K_{a},K_{c}}=3_{1,2}$ state of both H$_2^{16}$O and H$_2^{18}$O are observed in our spectrum, providing column densities of $N({\rm H_{2}^{16}O}|3_{1,2})=(21.1\pm0.84)\times10^{15}$~cm$^{-2}$ and $N({\rm H_{2}^{18}O}|3_{1,2})=(2.23\pm0.86)\times10^{15}$~cm$^{-2}$ (see Table \ref{tbl_results}). This state-specific analysis results in ${\rm H_2^{16}O/H_2^{18}O}=9.5$, much smaller than the oxygen isotopic ratio found in the solar system \citep[$^{16}$O/$^{18}$O $\approx500$;][and references therein]{asplund2009}. However, the H$_2^{16}$O $\nu_2=1-0$ $3_{2,1}$-$3_{1,2}$ transition is one of the strongest absorption features in our spectrum, suggesting that it traces a much smaller region than the optically thin H$_2^{18}$O $\nu_2=1-0$ $3_{2,1}$-$3_{1,2}$ transition. Rather than representing the real isotopic ratio then, this is another indicator of the optical depth effects discussed above. The best-fit temperature and H$_2^{16}$O column density can be used to estimate the state-specific column densities for H$_2^{16}$O corresponding to the two observed H$_2^{18}$O transitions for use in determining isotopic ratios. Using this method we find $N({\rm H_{2}^{16}O}|1_{0,1})=1.79\times10^{17}$~cm$^{-2}$ and $N({\rm H_{2}^{16}O}|3_{1,2})=2.71\times10^{17}$~cm$^{-2}$. Together with the measured H$_2^{18}$O state-specific column densities, these values give ${\rm H_2^{16}O/H_2^{18}O}\approx206$ and ${\rm H_2^{16}O/H_2^{18}O}\approx121$. Although still below the solar system isotopic ratio, these are well within the wide range of values found in other stellar atmospheres \citep{hinkle2016}.

\section{Summary}

We have observed H$_2$O absorption toward the massive protostar AFGL~2136~IRS~1 at high ($\sim3$~km~s$^{-1}$) spectral resolution in wavelength intervals near 2.5~$\mu$m, 6~$\mu$m, and 13~$\mu$m. H$_2$O absorption line profiles are consistent with those from other molecules that trace a hot gas component (e.g., NH$_3$, HCN, C$_2$H$_2$, HCl, highly excited CO), and do not match the profiles of sub-mm molecular emission lines that arise in the cooler, more extended envelope. Analysis of a rotation diagram constructed from state-specific column densities indicates an H$_2$O column density of $N({\rm H_2O})=(8.25\pm0.95)\times10^{18}$~cm$^{-2}$ at a temperature of $T=502\pm12$~K. It is clear, however, that an isothermal absorbing slab model in LTE does not adequately describe the relative intensities of all observed H$_2$O absorption lines. The most likely explanation for the H$_2$O line depths is that the absorbing gas is well mixed with the warm dust that serves as the ``background'' continuum source at 2.5--13~$\mu$m. Because the $\tau\approx1$ ``surface'' in this model is wavelength dependent and moves to shallower depths for stronger lines, the stronger lines are effectively probing a smaller region than the weaker lines. This is why the absorption features that should be the deepest are much shallower than expected, and also why fitting a line to only the upper envelope of points in the rotation diagram provides the best estimate of the total H$_2$O column density, although with the caveat that this still only provides the column density down to the $\tau\approx1$ ``surface''  for weak lines. 

Recent ALMA observations of AFGL~2136~IRS~1 have revealed a compact ($93\times71$~mas) disk that shows Keplerian rotation in the H$_2$O $\nu_2=1$--1, 5$_{5,0}$-6$_{4,3}$ transition at 232.687~GHz. The velocity at which the strongest H$_2$O absorption occurs in the infrared coincides with the H$_2$O emission velocity centroid along the disk minor axis. We conclude that the H$_2$O absorption observed in the infrared---as well as the absorption caused by other molecules that (1) have similar absorption profiles, and (2) indicate high rotational temperatures---arises within this circumstellar disk. Additionally, the warm dust in this disk is the dominant source of continuum emission at our observed wavelengths. Molecular absorption in the IR can thus serve as a powerful probe of accretion disks around massive protostars.

\vspace{1cm}
The authors thank the anonymous referee for providing suggestions to improve the quality of our paper. N.I. thanks M. Goto for providing VLT/CRIRES spectra of AFGL 2136 IRS 1 at 3.6~$\mu$m
so that HCl and H$_3^+$ line profiles could be included within this paper. M.J.R. and EXES observations are supported by NASA cooperative agreement NNX13AI85A. A.K. acknowledges support from the First TEAM grant of the Foundation for Polish Science No. POIR.04.04.00-00-5D21/18-00 and the Polish National Science Center grant 2016/21/D/ST9/01098. Support for this work was provided by the Polish National Agency for Academic Exchange through the project InterAPS.

Based on observations collected at the European Southern Observatory under ESO programmes  089.C-0321(B) and 091.C-0335(A). Based on observations made with the NASA/DLR Stratospheric Observatory for Infrared Astronomy (SOFIA). SOFIA is jointly operated by the Universities Space Research Association, Inc. (USRA), under NASA contract NNA17BF53C, and the Deutsches SOFIA Institut (DSI) under DLR contract 50 OK 0901 to the University of Stuttgart. [Financial support for this work was provided by NASA through award \#04-0120 issued by USRA.]  Based on observations obtained at the Gemini Observatory, which is operated by the Association of Universities for Research in Astronomy, Inc., under a cooperative agreement with the NSF on behalf of the Gemini partnership: the National Science Foundation (United States), National Research Council (Canada), CONICYT (Chile), Ministerio de Ciencia, Tecnolog\'{i}a e Innovaci\'{o}n Productiva (Argentina), Minist\'{e}rio da Ci\^{e}ncia, Tecnologia e Inova\c{c}\~{a}o (Brazil), and Korea Astronomy and Space Science Institute (Republic of Korea). Observations were obtained as part of program GN-2014B-Q-103.

\software{Astropy \citep{astropy2013}, CRIRES pipeline v2.3.3, IRAF \citep{tody1986,tody1993}, Matplotlib \citep{matplotlib2007}, Redux pipeline \citep{clarke2015}, Scipy \citep{scipy2019arxiv}, TEXES pipeline \citep{lacy2002}}

\bibliographystyle{aasjournal}
\bibliography{indy_master}


\begin{deluxetable}{cccccccc}
\tabletypesize{\scriptsize}
\tablecaption{Log of Observations of AFGL 2136 IRS 1 and Objects used as Telluric standards\label{tbl_obslog}}
\tablehead{\colhead{Target} & \colhead{Telescope} & \colhead{Instrument} & \colhead{Reference Wavelength} & \colhead{Date}  & \colhead{Altitude} & \colhead{Slit Width} & \colhead{Exposure Time}  \\
 & & & \colhead{(nm / cm$^{-1}$)\tablenotemark{a}} & & \colhead{(m)} & \colhead{(arcsec)} & \colhead{(s)}
 }
\startdata
AFGL 2136 IRS 1 & VLT UT1 & CRIRES & 2502.8 & 2012 Jul 6\tablenotemark{b} & 2635 & 0.2 & 3960 \\
AFGL 2136 IRS 1 & VLT UT1 & CRIRES & 2502.8 & 2013 Aug 11 & 2635 & 0.2 & 2340 \\
AFGL 2136 IRS 1 & VLT UT1 & CRIRES & 2480.0 & 2013 Aug 31 & 2635 & 0.2 & 2340 \\
AFGL 2136 IRS 1 & VLT UT1 & CRIRES & 2502.8 & 2013 Sep 16 & 2635 & 0.2 & 2160 \\
AFGL 2136 IRS 1 & VLT UT1 & CRIRES & 2480.0 & 2013 Sep 17 & 2635 & 0.2 & 2340 \\
AFGL 2136 IRS 1 & VLT UT1 & CRIRES & 2502.8 & 2013 Sep 19 & 2635 & 0.2 & 2340 \\
HR 6378 & VLT UT1 & CRIRES & 2502.8 & 2013 Aug 11 & 2635 & 0.2 & 300 \\
HR 6378 & VLT UT1 & CRIRES & 2480.0 & 2013 Aug 31 & 2635 & 0.2 & 840 \\
HR 6378 & VLT UT1 & CRIRES & 2502.8 & 2013 Sep 16 & 2635 & 0.2 & 300 \\
HR 6378 & VLT UT1 & CRIRES & 2480.0 & 2013 Sep 17 & 2635 & 0.2 & 300 \\
HR 6378 & VLT UT1 & CRIRES & 2502.8 & 2013 Sep 19 & 2635 & 0.2 & 300 \\
\hline
AFGL 2136 IRS 1 & SOFIA & EXES & 1485.90 & 2016 Mar 24 & 13410 & 1.84 & 650 \\
AFGL 2136 IRS 1 & SOFIA & EXES & 1640.15 & 2016 Mar 24 & 13102 & 1.84 & 720 \\
AFGL 2136 IRS 1 & SOFIA & EXES & 1748.25 & 2016 Mar 24 & 13100 & 1.84 & 480 \\
$\alpha$ CMa & SOFIA & EXES & 1485.90 & 2016 Mar 18 & 12814 & 1.84 & 884 \\
$\alpha$ CMa & SOFIA & EXES & 1640.15 & 2016 Mar 24 & 11593 & 1.84 & 840 \\
$\alpha$ CMa & SOFIA & EXES & 1747.58 & 2016 Mar 18 & 11280 & 1.84 & 416 \\
\hline
AFGL 2136 IRS 1 & Gemini N. & TEXES & 931.5 & 2014 Aug 11 & 4213 & 0.54 & 712 \\ 
AFGL 2136 IRS 1 & Gemini N. & TEXES & 856.0 & 2014 Aug 14 & 4213 & 0.54 & 210 \\
AFGL 2136 IRS 1 & Gemini N. & TEXES & 806.5 & 2014 Aug 17 & 4213 & 0.54 & 396 \\
AFGL 2136 IRS 1 & Gemini N. & TEXES & 768.5 & 2014 Aug 18 & 4213 & 0.54 & 406 \\ 
16 Psyche & Gemini N. & TEXES & 931.5 & 2014 Aug 11 & 4213 & 0.54 & 2372 \\ 
16 Psyche & Gemini N. & TEXES & 856.0 & 2014 Aug 14 & 4213 & 0.54 & 648 \\ 
16 Psyche & Gemini N. & TEXES & 806.5 & 2014 Aug 17 & 4213 & 0.54 & 1626 \\ 
16 Psyche & Gemini N. & TEXES & 768.5 & 2014 Aug 18 & 4213 & 0.54 & 1406 \\ 
\enddata
\tablenotetext{a}{Reference wavelengths for CRIRES observations are in units of nm, while those for EXES and TEXES observations are in cm$^{-1}$.}
\tablenotetext{b}{These observations have been presented in \citet{indriolo2013H2O}. There is no telluric standard associated with this observation as the target spectrum was divided by a model atmospheric spectrum.}
\end{deluxetable}
\normalsize


\clearpage
\startlongtable
\begin{deluxetable}{ccccccccc|c}
\tabletypesize{\scriptsize}
\tablecaption{Results from Unblended H$_2$O Absorption Lines\label{tbl_results}}
\tablehead{\colhead{$v'-v''$} & \colhead{$J_{K_{a},K_{c}}'-J_{K_{a},K_{c}}''$} & \colhead{Wavelength} & \colhead{$E_l$} & \colhead{$A_{ul}$} &\colhead{$g_u$}  & \colhead{$g_l$} & 
\colhead{$\int \tau dv$} & \colhead{ln$(N_l/g_l)$} & \colhead{$N_{\rm meas}/N_{\rm pred}$}  \\
 & & \colhead{($\mu$m)} & \colhead{(K)} & \colhead{(s$^{-1}$)} & \colhead{} & \colhead{} & \colhead{(km~s$^{-1}$)} & \colhead{} &  
 }
 \startdata
$\nu_3=1-0$ &  $8_{3,5}$-$7_{1,6}$ &    2.446108 &  1013.2 &  0.979 &  51 & 45 &   0.79$\pm$0.15 &  35.53$\pm$0.19  &  0.96  \\
$\nu_3=1-0$ &  $5_{5,1}$-$4_{3,2}$ &    2.447276 &   550.4 &  0.082 &  33 & 27 &   0.18$\pm$0.16 &  36.95$\pm$0.90  &  1.58  \\
$\nu_3=1-0$ &  $7_{2,5}$-$6_{0,6}$ &    2.450228 &   642.7 &  0.321 &  15 & 13 &   0.20$\pm$0.15 &  36.52$\pm$0.75  &  1.23  \\
$\nu_3=1-0$ &  $8_{4,4}$-$7_{2,5}$ &    2.451345 &  1125.7 &  1.480 &  51 & 45 &   0.87$\pm$0.15 &  35.22$\pm$0.17  &  0.88  \\
$\nu_3=1-0$ &  $14_{5,9}$-$13_{5,8}$ &    2.453230 &  3783.0 &  20.510 &  87 & 81 &   0.19$\pm$0.17 &  30.50$\pm$0.92  &  1.56  \\
$\nu_3=1-0$ &  $6_{4,3}$-$5_{2,4}$ &    2.454677 &   598.8 &  0.339 &  13 & 11 &   0.13$\pm$0.14 &  36.16$\pm$1.04  &  0.78  \\
$\nu_3=1-0$ &  $14_{4,10}$-$13_{4,9}$ &    2.461582 &  3645.6 &  25.020 &  87 & 81 &   0.22$\pm$0.16 &  30.46$\pm$0.72  &  1.13  \\
$\nu_1=1-0$ &  $8_{4,5}$-$7_{1,6}$ &    2.467807 &  1013.2 &  0.194 &  51 & 45 &   0.27$\pm$0.16 &  36.07$\pm$0.58  &  1.64  \\
$\nu_1=1-0$ &  $12_{6,7}$-$11_{5,6}$ &    2.468616 &  2876.1 &  6.393 &  75 & 69 &   0.22$\pm$0.17 &  31.96$\pm$0.76  &  1.10  \\
$\nu_3=1-0$ &  $13_{4,9}$-$12_{4,8}$ &    2.468911 &  3173.4 &  25.930 &  27 & 25 &   0.15$\pm$0.15 &  31.24$\pm$0.98  &  0.97  \\
$\nu_3=1-0$ &  $7_{3,4}$-$6_{1,5}$ &    2.472935 &   781.1 &  1.202 &  15 & 13 &   0.51$\pm$0.15 &  36.09$\pm$0.30  &  1.05  \\
$\nu_1=1-0$ &  $11_{6,5}$-$10_{5,6}$ &    2.473202 &  2472.8 &  2.726 &  69 & 63 &   0.18$\pm$0.16 &  32.69$\pm$0.87  &  1.02  \\
$\nu_3=1-0$ &  $14_{6,8}$-$13_{6,7}$ &    2.473837 &  3965.9 &  16.270 &  87 & 81 &   0.11$\pm$0.17 &  30.16$\pm$1.62  &  1.60  \\
$\nu_1=1-0$ &  $11_{5,6}$-$10_{4,7}$ &    2.475249 &  2275.2 &  2.115 &  69 & 63 &   0.22$\pm$0.20 &  33.17$\pm$0.89  &  1.12  \\
$\nu_3=1-0$ &  $14_{3,11}$-$13_{3,10}$ &    2.477481 &  3474.2 &  24.950 &  87 & 81 &   0.32$\pm$0.16 &  30.82$\pm$0.51  &  1.16  \\
$\nu_3=1-0$ &  $12_{4,8}$-$11_{4,7}$ &    2.478598 &  2732.2 &  26.170 &  75 & 69 &   0.95$\pm$0.15 &  32.01$\pm$0.16  &  0.87  \\
$\nu_1=1-0$ &  $9_{3,6}$-$8_{2,7}$ &    2.479768 &  1274.2 &  0.170 &  57 & 51 &   0.17$\pm$0.17 &  35.57$\pm$1.05  &  1.68  \\
$\nu_1=1-0$ &  $5_{4,1}$-$4_{1,4}$ &    2.480526 &   323.5 &  0.049 &  33 & 27 &   0.14$\pm$0.17 &  37.17$\pm$1.23  &  1.24  \\
$\nu_3=1-0$ &  $13_{3,10}$-$12_{3,9}$ &    2.483376 &  3029.9 &  25.500 &  27 & 25 &   0.20$\pm$0.15 &  31.48$\pm$0.79  &  0.92  \\
$\nu_3=1-0$ &  $6_{2,4}$-$5_{0,5}$ &    2.484255 &   468.1 &  0.494 &  39 & 33 &   1.05$\pm$0.14 &  36.73$\pm$0.14  &  1.08  \\
$\nu_1=1-0$ &  $10_{6,4}$-$9_{5,5}$ &    2.486180 &  2122.2 &  2.179 &  21 & 19 &   0.12$\pm$0.16 &  33.66$\pm$1.40  &  1.34  \\
$\nu_2=2-0$ &  $15_{3,12}$-$14_{2,13}$ &    2.488795 &  3349.3 &  8.645 &  93 & 87 &   0.18$\pm$0.17 &  31.23$\pm$0.91  &  1.37  \\
$\nu_3=1-0$ &  $11_{4,7}$-$10_{4,6}$ &    2.490420 &  2325.7 &  24.690 &  23 & 21 &   0.70$\pm$0.17 &  32.93$\pm$0.24  &  0.97  \\
$\nu_3=1-0$ &  $13_{4,10}$-$12_{4,9}$ &    2.491232 &  3057.3 &  24.010 &  81 & 75 &   0.55$\pm$0.16 &  31.46$\pm$0.28  &  0.96  \\
$\nu_3=1-0$ &  $4_{4,0}$-$3_{2,1}$ &    2.492091 &   305.2 &  0.127 &  27 & 21 &   0.27$\pm$0.14 &  37.10$\pm$0.51  &  1.12  \\
$\nu_3=1-0$ &  $6_{3,3}$-$5_{1,4}$ &    2.494654 &   574.7 &  0.922 &  39 & 33 &   1.27$\pm$0.14 &  36.28$\pm$0.11  &  0.85  \\
$\nu_3=1-0$ &  $11_{3,8}$-$10_{3,7}$ &    2.496575 &  2213.0 &  27.470 &  23 & 21 &   0.67$\pm$0.15 &  32.77$\pm$0.22  &  0.66  \\
$\nu_3=1-0$ &  $13_{2,11}$-$12_{2,10}$ &    2.497063 &  2820.3 &  27.190 &  27 & 25 &   0.22$\pm$0.13 &  31.52$\pm$0.61  &  0.63  \\
$\nu_3=1-0$ &  $12_{2,10}$-$11_{2,9}$ &    2.504591 &  2432.5 &  27.810 &  75 & 69 &   1.60$\pm$0.15 &  32.44$\pm$0.09  &  0.73  \\
$\nu_3=1-0$ &  $11_{4,8}$-$10_{4,7}$ &    2.511261 &  2275.2 &  23.560 &  69 & 63 &   1.81$\pm$0.15 &  32.80$\pm$0.08  &  0.77  \\
$\nu_3=1-0$ &  $11_{2,9}$-$10_{2,8}$ &    2.512036 &  2068.9 &  27.820 &  23 & 21 &   1.17$\pm$0.15 &  33.30$\pm$0.13  &  0.84  \\
$\nu_3=1-0$ &  $11_{5,7}$-$10_{5,6}$ &    2.512708 &  2472.8 &  19.930 &  69 & 63 &   1.27$\pm$0.16 &  32.62$\pm$0.12  &  0.95  \\
$\nu_3=1-0$ &  $9_{3,6}$-$8_{3,5}$ &    2.517545 &  1510.9 &  28.400 &  19 & 17 &   2.33$\pm$0.14 &  34.15$\pm$0.06  &  0.65  \\
$\nu_3=1-0$ &  $11_{6,6}$-$10_{6,5}$ &    2.518460 &  2697.7 &  17.010 &  69 & 63 &   0.46$\pm$0.15 &  31.76$\pm$0.33  &  0.63  \\
$\nu_3=1-0$ &  $10_{2,8}$-$9_{2,7}$ &    2.519438 &  1729.3 &  28.430 &  63 & 57 &   3.95$\pm$0.13 &  33.48$\pm$0.03  &  0.51  \\
$\nu_2=1-0$ &  $7_{4,4}$-$7_{3,5}$ &    5.705113 &  1175.0 &  2.981 &  15 & 15 &   3.97$\pm$0.50 &  34.72$\pm$0.12  &  0.59  \\
$\nu_2=1-0$ &  $8_{1,8}$-$7_{0,7}$ &    5.709642 &   843.5 &  11.840 &  51 & 45 &  15.25$\pm$1.18 &  33.46$\pm$0.08  &  0.09  \\
$\nu_2=1-0$ &  $11_{3,8}$-$11_{2,9}$ &    5.713061 &  2432.5 &  3.049 &  69 & 69 &   1.71$\pm$0.47 &  32.33$\pm$0.27  &  0.66  \\
$\nu_2=1-0$ &  $6_{4,3}$-$6_{3,4}$ &    5.716237 &   933.7 &  2.772 &  39 & 39 &   8.11$\pm$0.52 &  34.55$\pm$0.06  &  0.31  \\
$\nu_2=1-0$ &  $5_{2,4}$-$4_{1,3}$ &    5.718679 &   396.4 &  5.157 &  11 & 9 &   9.94$\pm$6.93 &  35.39$\pm$0.70  &  0.24  \\
$\nu_2=2-1$ &  $10_{1,10}$-$9_{0,9}$ &    5.720222 &  3614.6 &  25.810 &  63 & 57 &   2.04$\pm$0.47 &  30.46$\pm$0.23  &  1.06  \\
$\nu_2=1-0$ &  $5_{4,2}$-$5_{3,3}$ &    5.721717 &   725.1 &  2.390 &  11 & 11 &   5.76$\pm$0.50 &  35.62$\pm$0.09  &  0.59  \\
$\nu_2=1-0$ &  $8_{1,7}$-$8_{0,8}$ &    5.726424 &  1070.5 &  1.397 &  17 & 17 &   2.48$\pm$0.46 &  34.87$\pm$0.19  &  0.56  \\
$\nu_2=1-0$ &  $4_{4,0}$-$4_{3,1}$ &    5.728111 &   552.3 &  1.617 &  9 & 9 &   4.28$\pm$0.48 &  35.91$\pm$0.11  &  0.56  \\
$\nu_2=1-0$ &  $8_{3,6}$-$8_{2,7}$ &    5.731998 &  1274.2 &  2.692 &  51 & 51 &   6.38$\pm$0.53 &  34.06$\pm$0.08  &  0.37  \\
$\nu_2=1-0$ &  $3_{2,1}$-$3_{1,2}$ &    6.075447 &   249.4 &  6.715 &  21 & 21 &  12.62$\pm$0.50 &  34.54$\pm$0.04  &  0.08  \\
$\nu_2=2-1$ &  $5_{1,4}$-$5_{0,5}$ &    6.080180 &  2763.5 &  6.406 &  33 & 33 &   1.78$\pm$0.58 &  32.18$\pm$0.33  &  1.09  \\
$\nu_2=1-0$ &  $5_{3,2}$-$4_{4,1}$ &    6.088700 &   702.3 &  0.351 &  33 & 27 &   3.48$\pm$0.40 &  35.75$\pm$0.12  &  0.64  \\
$\nu_2=1-0$ &  $5_{2,4}$-$4_{3,1}$ &    6.096408 &   552.3 &  0.733 &  11 & 9 &   3.33$\pm$0.34 &  36.06$\pm$0.10  &  0.65  \\
$\nu_2=2-1$ &  $2_{2,1}$-$2_{1,2}$ &    6.100969 &  2412.9 &  5.916 &  15 & 15 &   0.84$\pm$0.32 &  32.28$\pm$0.38  &  0.60  \\
$\nu_2=1-0$ &  $5_{3,3}$-$4_{4,0}$ &    6.103987 &   702.3 &  0.343 &  11 & 9 &   1.72$\pm$0.38 &  36.16$\pm$0.22  &  0.96  \\
$\nu_2=1-0$ &  $4_{2,2}$-$3_{3,1}$ &    6.106193 &   410.4 &  0.529 &  9 & 7 &   2.73$\pm$0.31 &  36.38$\pm$0.11  &  0.67  \\
$\nu_2=1-0$ &  $3_{1,2}$-$2_{2,1}$ &    6.106826 &   194.1 &  1.123 &  21 & 15 &   9.60$\pm$0.72 &  36.04$\pm$0.08  &  0.31  \\
$\nu_2=2-1$ &  $4_{0,4}$-$3_{1,3}$ &    6.113164 &  2502.7 &  15.800 &  9 & 7 &   1.44$\pm$0.36 &  32.34$\pm$0.25  &  0.77  \\
$\nu_2=1-0$ &  $1_{1,1}$-$0_{0,0}$ &    6.116331 &     0.0 &  7.457 &  3 & 1 &  12.55$\pm$0.66 &  36.36$\pm$0.05  &  0.29  \\
$\nu_2=1-0$ &  $10_{3,8}$-$10_{4,7}$ &    6.705153 &  2275.2 &  9.459 &  63 & 63 &   4.03$\pm$0.27 &  31.66$\pm$0.07  &  0.25  \\
$\nu_2=1-0$ &  $4_{0,4}$-$5_{1,5}$ &    6.707693 &   469.9 &  8.449 &  9 & 11 &   5.46$\pm$0.24 &  34.03$\pm$0.04  &  0.07  \\
$\nu_2=2-1$ &  $4_{2,2}$-$4_{3,1}$ &    6.711310 &  2886.1 &  14.320 &  9 & 9 &   0.89$\pm$0.25 &  31.69$\pm$0.28  &  0.85  \\
$\nu_2=1-0$ &  $6_{1,6}$-$6_{2,5}$ &    6.712122 &   795.5 &  4.825 &  39 & 39 &   6.10$\pm$0.33 &  33.23$\pm$0.05  &  0.06  \\
$\nu_2=1-0$ &  $6_{3,4}$-$7_{2,5}$ &    6.714553 &  1125.7 &  1.928 &  39 & 45 &   5.91$\pm$0.26 &  34.11$\pm$0.04  &  0.29  \\
$\nu_2=1-0$ &  $8_{2,7}$-$8_{3,6}$ &    6.715692 &  1447.6 &  7.602 &  51 & 51 &   6.52$\pm$0.25 &  32.57$\pm$0.04  &  0.12  \\
$\nu_2=1-0$ &  $13_{4,10}$-$13_{5,9}$ &    6.722649 &  3721.4 &  11.270 &  27 & 27 &   0.41$\pm$0.23 &  30.04$\pm$0.56  &  0.87  \\
$\nu_2=1-0$ &  $2_{1,1}$-$3_{2,2}$ &    6.723374 &   296.8 &  8.223 &  5 & 7 &   5.22$\pm$0.58 &  34.59$\pm$0.11  &  0.09  \\
$\nu_2=2-1$ &  $7_{1,6}$-$7_{2,5}$ &    6.726109 &  3442.4 &  20.140 &  45 & 45 &   1.59$\pm$0.27 &  30.30$\pm$0.17  &  0.65  \\
$\nu_2=2-1$ &  $3_{2,1}$-$3_{3,0}$ &    6.733401 &  2744.6 &  9.039 &  21 & 21 &   1.21$\pm$0.26 &  31.59$\pm$0.21  &  0.59  \\
$\nu_2=1-0$ &  $10_{5,6}$-$11_{4,7}$ &    6.737377 &  2732.2 &  0.804 &  63 & 69 &   0.34$\pm$0.23 &  31.65$\pm$0.67  &  0.60  \\
$\nu_2=2-1$ &  $7_{3,4}$-$7_{4,3}$ &    6.753573 &  3700.7 &  17.650 &  45 & 45 &   1.07$\pm$0.26 &  30.03$\pm$0.24  &  0.82  \\
 &  $16_{3,14}$-$15_{0,15}$ &   11.724550 &  3393.1 &  1.682 &  99 & 93 &   0.26$\pm$0.03 &  28.52$\pm$0.13  &  0.10  \\
 &  $16_{4,13}$-$15_{1,14}$ &   12.375657 &  3785.8 &  4.211 &  99 & 93 &   0.99$\pm$0.05 &  28.78$\pm$0.05  &  0.28  \\
 &  $16_{3,13}$-$15_{2,14}$ &   12.407046 &  3785.8 &  4.217 &  33 & 31 &   0.35$\pm$0.06 &  28.82$\pm$0.17  &  0.29  \\
 &  $16_{5,12}$-$15_{2,13}$ &   13.033327 &  4132.6 &  6.761 &  99 & 93 &   0.64$\pm$0.06 &  27.71$\pm$0.09  &  0.19  \\
\hline
\multicolumn{10}{c}{H$_2^{18}$O} \\
\hline
$\nu_2=1-0$ &  $2_{1,2}$-$1_{0,1}$ &    6.074293 &    34.2 &  6.533 &  15 & 9 &   0.84$\pm$0.34 &  32.20$\pm$0.40  & ...  \\
$\nu_2=1-0$ &  $3_{2,1}$-$3_{1,2}$ &    6.103487 &   248.7 &  6.404 &  21 & 21 &   1.29$\pm$0.50 &  32.30$\pm$0.39  & ...  \\
\enddata
\tablecomments{Transition properties for all unblended lines fit during our analysis are shown here. Integrated optical depths ($\int\tau dv$) are over the entire absorption feature (i.e., the sum of the two components). Individual state column densities assuming optically thin absorption are presented as $\ln(N_l/g_l)$. These values should be considered with extreme caution given the discussion presented in this paper. The last column gives the ratio between the measured and predicted state-specific column densities. Predicted column densities assume an isothermal absorbing slab in LTE with the temperature and total H$_2$O column density determined from CRIRES observations alone (see Table \ref{tbl_rotdiagcomp}). Smaller values correspond to more ``under-populated'' states.}
\end{deluxetable}
\normalsize

\clearpage
\begin{deluxetable}{lcc}
\tablecaption{Rotation Diagram Analysis Comparison \label{tbl_rotdiagcomp}}
\tablehead{\colhead{Data Sub-Sample} & \colhead{Temperature} & \colhead{$N({\rm H_2O})$} \\
 & \colhead{(K)} & \colhead{(10$^{18}$ cm$^{-2}$)}}
\startdata
All Data 2.44--13.1~$\mu$m & $545\pm24$ & $4.13\pm0.80$ \\
CRIRES 2.44--2.52~$\mu$m & $502\pm12$\tablenotemark{a} & $8.25\pm0.95$\tablenotemark{a} \\
EXES 5.7--6.76~$\mu$m & $628\pm44$ & $2.48\pm0.60$ \\
EXES 5.7--5.74~$\mu$m & $627\pm81$ & $2.94\pm1.15$ \\
EXES 6.07--6.12~$\mu$m& $626\pm85$ & $3.85\pm1.48$ \\
EXES 6.7--6.76~$\mu$m & $812\pm63$ & $1.05\pm0.28$ \\
\hline
EXES 5.7--6.76~$\mu$m, $\tau_0<0.29$ & $519\pm19$ & $5.13\pm0.94$ \\
\enddata
\tablecomments{Temperatures and total H$_2$O column densities are inferred from linear fits to rotation diagrams made from different sub-sets of individual state column densities determined from unblended absorption lines.}
\tablenotetext{a}{Adopted values}
\end{deluxetable}


\clearpage

\clearpage

\clearpage

\clearpage

\clearpage

\clearpage

\clearpage

\clearpage

\clearpage

\clearpage

\end{document}